\documentclass[letterpaper,titlepage,11pt]{article}
\usepackage{latexsym,amssymb,amstext,amsmath}
\usepackage{slashed}
\usepackage{mathrsfs}
\usepackage{color}
\usepackage{mathtools}
\usepackage{enumerate}
\usepackage{graphicx}
\usepackage{tikz}
\usetikzlibrary{shapes,arrows}
\usepackage{etex}
\usepackage{latexsym}
\usepackage{cite}
\usepackage{accents}

\allowdisplaybreaks

\usepackage[
colorlinks=true,
linkcolor=blue,
urlcolor=red,
filecolor=green,
citecolor=red,
pdfstartview=FitV,
pdftitle={},
pdfauthor={Mehmet Ozkan},
pdfsubject={},
pdfkeywords={},
pdfpagemode=None,
bookmarksopen=true
]{hyperref}

\newcommand{\bea}{\setlength\arraycolsep{2pt} \begin{eqnarray}}
	\newcommand{\eea}{\end{eqnarray}}
\newcommand{\nn}{\nonumber}

\usepackage{hyperref}

\def\rmi{{\rm i}}
\newsavebox{\uuunit}
\sbox{\uuunit}
{\setlength{\unitlength}{0.825em}
	\begin{picture}(0.6,0.7)
		\thinlines
		\put(0,0){\line(1,0){0.5}}
		\put(0.15,0){\line(0,1){0.7}}
		\put(0.35,0){\line(0,1){0.8}}
		\multiput(0.3,0.8)(-0.04,-0.02){12}{\rule{0.5pt}{0.5pt}}
		\end {picture}}

	\def\be{\begin{equation}}
		\def\ee{\end{equation}}
	\def\ba{\begin{array}}
		\def\ea{\end{array}}
	\def\bea{\begin{eqnarray}}
		\def\eea{\end{eqnarray}}
	\def\bd{\begin{displaymath}}
		\def\ed{\end{displaymath}}
	
	\def\nn{\nonumber}
	
	\def\a{\alpha}
	\def\b{\beta}
	\def\g{\gamma}
	\def\G{\Gamma}
	\def\d{\delta}
	
	\def\e{\epsilon}
	
	\def\f{\phi}
	\def\vf{\varphi}
	
	\def\p{\psi}

	\def\l{\lambda}
	\def\L{\Lambda}
	\def\m{\mu}
	\def\n{\nu}

	\def\t{\tau}

	\def\x{\xi}
	
	\def\nn{\nonumber}
	\def\cD{\mathcal{D}}
	\def\cN{\mathcal{N}}
	
	\def\cL{\mathcal{L}}

	\makeatletter
	\@addtoreset{equation}{section}
	\makeatother
	
\begin{document}
		%
		\begin{titlepage}
			
			\bigskip
			
			\begin{center}
				{\LARGE\bfseries  Lie Algebra Expansions, Non-Relativistic Matter Multiplets and Actions}
				\\[10mm]
				\textbf{Oguzhan Kasikci and Mehmet Ozkan}\\[5mm]
				\vskip 25pt

				{\em  \hskip -.1truecm Department of Physics, Istanbul Technical University,  \\
					Maslak 34469 Istanbul, Turkey  \vskip 10pt }

				{email: {\tt kasikcio@itu.edu.tr, ozkanmehm@itu.edu.tr}}
				
			\end{center}
			
			\vspace{3ex}

			\begin{center}
				{\bfseries Abstract}
			\end{center}
			\begin{quotation} \noindent
				
				We discuss a general methodology to provide rigid, off-shell matter multiplets and actions for recently constructed non-relativistic superalgebras. The technique is based on the Lie algebra expansion, which, in the context of supersymmetry, has so far been used to obtain non-relativistic on-shell supergravity models. We first explain how the Lie algebra expansion can be implemented to generate off-shell rigid multiplets on a flat background by developing an auxiliary framework where only the lowest order behavior of the spatial and temporal vielbein is relevant. We then provide explicit examples for the field content, the transformation rules as well as action principles for certain multiplets in three and four dimensions. 
				
			\end{quotation}
			
			\vfill
			
			\flushleft{\today}
		\end{titlepage}
		\setcounter{page}{1}
		\tableofcontents
		
		\newpage

		\section{Introduction}{\label{Intro}}
		\paragraph{}
		Recent years have witnessed a growing interest in non-relativistic gravity from different directions including string theory \cite{Gomis:2000bd,Andringa:2012uz,Harmark:2017rpg,Kluson:2018egd,Kluson:2018grx,Harmark:2018cdl,Bergshoeff:2018yvt,Bergshoeff:2018vfn,Gomis:2019zyu,Bergshoeff:2019pij,Gallegos:2019icg,Harmark:2019upf,Kluson:2019ifd,Kluson:2019xuo,Blair:2019qwi,Kluson:2020aoq,Gomis:2020izd, Gomis:2020fui, Bergshoeff:2021bmc, Bergshoeff:2021tfn}, (super)gravity \cite{Papageorgiou:2009zc,Andringa:2013mma,Bergshoeff:2015uaa,Bergshoeff:2015ija,Bergshoeff:2016lwr,Hartong:2016yrf,Hartong:2017bwq,Bergshoeff:2017dqq,Matulich:2019cdo,Ozdemir:2019orp,Aviles:2019xed,Chernyavsky:2019hyp,Penafiel:2019czp,Ozdemir:2019tby,Bergshoeff:2020fiz,Hartong:2015zia,Afshar:2015aku,Abedini:2019voz,Kasikci:2020qsj,Concha:2020ebl,Concha:2020sjt,Concha:2020tqx,Concha:2021jos}, group theory \cite{Andringa:2010it,Figueroa-OFarrill:2017sfs,Figueroa-OFarrill:2018ilb,Bergshoeff:2019ctr,Gomis:2019fdh,Figueroa-OFarrill:2019sex,deAzcarraga:2019mdn,Gomis:2019sqv,Fontanella:2020eje,Concha:2020eam} and condensed matter physics \cite{Christensen:2013lma,Christensen:2013rfa,Hartong:2014oma,Salgado-Rebolledo:2021wtf}.  Although the original attempts were limited to Newton–Cartan (NC) geometry to propose a geometric formulation for the field equations of Newtonian gravity \cite{Cartan1,Cartan2}, it is now evident that one must go beyond the standard NC-geometry to formulate an action principle for Newtonian gravity, thanks to the recent work on large-$c$ expansion of the Einstein's general relativity \cite{VandenBleeken:2017rij,Hansen:2019pkl,VandenBleeken:2019gqa,Hansen:2019svu,Hansen:2019vqf,Ergen:2020yop}. This expansion also suggests that one must go beyond the Bargmann algebra as the algebra of the underlying symmetries of Newtonian gravity which then brings about the question that how does these new currents couple to matter. This question was answered in the foundational work \cite{Hansen:2020pqs} by the large-$c$ expansion of the matter fields. However, as the approach of \cite{Hansen:2020pqs} to matter couplings of non-relativistic theories is based on the metric formulation, it is natural to ask if we can describe the matter couplings in a gauge theory formulation. This question is particularly important for the non-relativistic supersymmetric theories where the gauge theory formulation is essential. 
		
		From a gauge theory perspective, there are essentially two different methodologies to generate non-relativistic (super)algebras that go beyond the Galilei or Bargmann superalgebra
		\begin{itemize}
			\item{{Lie Algebra (or Semi-Group) Expansion:}  The main idea of this methodology is to start from a smaller algebra and generate larger ones by the consistent truncation of infinite series expansion of Maurer-Cartan one-forms or the generators corresponding to the Maurer-Cartan one-forms \cite{Hatsuda:2001pp,deAzcarraga:2002xi,Izaurieta:2006zz,deAzcarraga:2007et}. Examples of bosonic and supersymmetric models can be found in \cite{Bergshoeff:2019ctr,deAzcarraga:2019mdn,Gomis:2019fdh,Ozdemir:2019tby,Penafiel:2019czp,Gomis:2019sqv,Concha:2019lhn,Concha:2019dqs,Gomis:2019nih,Kasikci:2020qsj,Concha:2020sjt,Fontanella:2020eje,Concha:2020ebl,Concha:2020tqx,Concha:2020eam}.}
			\item{{Coadjoint Construction:} In the coadjoint construction, one starts with a relativistic algebra and expands it to a certain order by using Lie algebra or S-expansion. The resulting algebra is then contracted, or improved by $U(1)$ generators and then contracted, depending on the non-relativistic algebra under consideration. See \cite{Bergshoeff:2020fiz} for bosonic examples.}
		\end{itemize}
		Although formally these methodologies may be applied to any relativistic or non-relativistic algebra, the most interesting ones arise when they are applied to the Poincar\'e algebra, either directly \cite{Bergshoeff:2020fiz}  or to its spacetime decomposition \cite{Bergshoeff:2019ctr}.  It is, therefore, an important question whether we can utilize these methods to construct matter multiplets of the extended algebras that we obtain from various expansions of the Poincar\'e superalgebra. In particular, given the fact that non-relativistic supersymmetry is mostly demanded in obtaining non-relativistic SUSY-QFTs on curved background where the off-shell formulation of the supergravity and the matter multiplets is essential \cite{Bergshoeff:2020baa}, it would be much desired to use these methodologies to find off-shell supermultiplets for the non-relativistic extended algebras.
		
		At first sight, this may look like a straightforward task, that is, one can simply take an off-shell multiplet and consider a series expansion of its fields. However, there are various subtleties for such a direct application. First of all, both of the expansion procedures that we described here are based on the gauging of the (super)algebra. This point is one of the most crucial to these constructions so let us be more precise. Consider an algebra $\mathfrak{g}$ with a set of generators $\{X_i, Y_\alpha\}$ where $X_i$ represent the even subset of the generators, which we denote by $V_0$, while $Y_\alpha$ represents the odd subset, $V_1$. Here, the even/odd splitting of the generators are based on the following rules on the commutation relations \cite{Hatsuda:2001pp,deAzcarraga:2002xi,deAzcarraga:2007et}
		\begin{eqnarray}\label{EvenOdd}
			[V_0, V_0] \subset V_0 \,, \qquad [V_0, V_1] \subset V_1 \,, \qquad [V_1, V_1] \subset V_0 \,.
		\end{eqnarray}
		Based on this splitting of the algebra, we may assign a gauge field to each generator, i.e., 
		\begin{eqnarray}
			A = A^i X_i  +  A^\alpha Y_\alpha \,,
		\end{eqnarray}	
		where $A = A_\mu dx^\mu$. This splitting of the gauge fields can be followed by the expansion
		\begin{eqnarray} \label{ExpansionGaugeFields}
			A^i = \sum_{n=0}^{N_0} \lambda^{2n} A_{(2n)}^{i} \,, \qquad A^\alpha = \sum_{n=0}^{N_1} \lambda^{2n+1} A_{(2n+1)}^{\alpha} \,.
		\end{eqnarray}
		This sum essentially goes all the way to infinity but we may terminate the sum at ${g} = (N_0, N_1)$ order, where $N_0$ represents the truncation order of even gauge fields while $N_1$ represents the truncation of the odd gauge fields. The consistency of this truncation is that either $N_0 = N_1 + 1$ or $N_0 = N_1$ must be satisfied. Based on this expansion of the gauge fields, one can expand the group-theoretical curvatures and read off the structure constants from the expanded curvatures to establish the larger algebra. This technical note for the expansion clearly shows that it is a local procedure, hence one would need an off-shell non-relativistic supergravity to build upon the off-shell local multiplets. On the other hand, the expansion of the gauge fields that are associated with the Poincar\'e superalgebra definitely does not give rise to an off-shell formulation of supergravity, but the superalgebra closes on the gauge fields up to curvature constraints and field equations. One may also be tempted to formulate a non-relativistic off-shell supergravity by expanding a relativistic one as a first step. However, as explained in \cite{Bergshoeff:2015uaa}, even the contraction of an off-shell Poincar\'e supergravity, which corresponds to the lowest order expansion of fields, is quite non-trivial as the number of auxiliary fields in a non-relativistic case may not be as many as what one has in the relativistic case. Furthermore, it is not guaranteed that the expansion of the composite relativistic expressions, such as the spin connection, does not spoil the covariance of the covariant objects in the expansion procedure. 
	
		Although local off-shell supermultiplets may be inaccessible from an expansion point of view, it might be possible to obtain rigid off-shell supermultiplets by applying the following prescription. Consider a field $\Phi$ which transforms non-trivially under the action of the generators of the  algebra $\mathfrak{g}$ that represents either the Poincar\'e superalgebra itself or its spacetime decomposition, i.e.
		\begin{eqnarray}\label{Commutator}
			[\mathcal D_\mu, \mathcal D_\nu] \Phi = - \delta(R_{\mu\nu} (X_i)) \Phi -  \delta(R_{\mu\nu} (Y_\alpha)) \Phi \,.
		\end{eqnarray}
		Here, $ \delta (R_{\mu\nu})$ means that one calculates the transformation of the field $\Phi$ with respect to all gauge symmetries generated by $\{X_i, Y_\alpha\}$ and replace the relevant transformation parameters with the group-theoretical curvatures. Although the expansion procedure must be realized with full local transformation rules, taking the rigid limit on a flat background indicates that in the end, we need to set the vielbein to $e_\m{}^a = \d_\m{}^a$ (or set the spatial vielbein $e_\m{}^a = \d_{\m}{}^a$ and temporal vielbein $\t_\m = \d_\m^0$) and all other fields of the off-shell Poincar\'e multiplet to zero. Consequently, we may introduce an auxiliary framework where we start with the transformation rules of an off-shell, rigid matter multiplet but introduce a vielbein (or both the spatial and the temporal vielbein in the case of spacetime decomposed algebra) to relate the flat indices on derivatives to curved indices. The reason for that is because the Lie algebra expansion is established by the expansion of the gauge fields and curvatures that carry curved indices. Hence the expansion can only be realized after all  flat indices are converted to curved indices by use of vielbein or inverse vielbein. Note that in the case of spacetime decompose algebra, this would mean that the use of temporal vielbein ($\tau_\mu$) or its inverse $(\tau^\mu)$ is needed to be used to convert temporal index to a curved index. Once this step is done, the commutation relation \eqref{Commutator} determines the expansion character of the field $\Phi$. We can then expand the fields $\Phi$ as well as the curvatures in accordance with the expansion of the gauge fields \eqref{ExpansionGaugeFields}. However, it is important to keep in mind that we only use the first-order scaling character of the vielbein and the temporal vielbein as their expansion introduces extra gauge fields which are truncated in the rigid limit. Finally, we can set $e_\m{}^a = \d_{\m}{}^a$ and $\t_\m = \d_\m^0$ and read off the transformation rules of the expanded fields for each $\l$-order term in Eq. \eqref{Commutator} to obtain the rigid supersymmetric multiplets of the expanded algebras. 
		
		Before we proceed to the actual construction of the matter multiplets of expanded superalgebras, let us briefly discuss the difference between the methodology of \cite{Hansen:2020pqs} which is based on the metric construction, and a procedure that suits a gauge theory construction. In a metric formulation that is based on a $1/c$ expansion procedure, all the Galilean building blocks are expanded in even-powers of $1/c$. Hence it is natural to expand fields in even powers of $1/c$ along with the transformation parameters to find the local transformation rules. On the other hand, in gauge theory construction, it is essential that some of the gauge fields, or generators, are expanded in odd-powers of $\lambda$. In this case, the expansion character of the field is not as clear as in the $1/c$ expansion. Another question that comes to mind is whether one can understand the operator structure of the generators in the larger algebra starting from the smaller algebra as they are related to each other by means of expansion. This can be achieved by expanding the space-time coordinates based on the expansion character of the generator that generates translations for that coordinate  \cite{Barducci:2020blv}. In that case, the original dimension of the spacetime increases thus one needs to develop a dimensional reduction scheme to find the desired representations of the expanded algebra in the same number of dimensions that one starts with. For a supersymmetric theory, this increases the workload tremendously since one now needs to construct the supersymmetric representations of the expanded superalgebra, which is already a non-Lorentzian algebra with a presumably arbitrary number of time coordinates due to the expansion of the original time coordinate. It is also necessary to find a way for the dimensional reduction of this representation to the spacetime dimension that one starts with. Thus, although tempting in terms of the operator definitions of the generators, we do not follow this path for the bosonic generators but directly construct their actions on fields utilizing the expansion of \eqref{Commutator}. This gives the operator definitions of generators a non-linear matrix character in the sense that the matrix part of a generator becomes a matrix of operators. For the fermionic generators, however, it is better to expand the fermionic coordinates of superspace as more fermionic generators already imply the increasing number of supercharges, thus the fermionic coordinates of the superspace. This is also the main procedure that determines the expansion character of the superfield, allowing us to find the supersymmetric representations of the expanded, larger algebras.  
		
		This paper is organized as follows. In Section \ref{Section2}, we discuss the basics of the expansion of an off-shell matter multiplet. As mentioned, the consistency of the truncation for the Lie algebra expansion requires that either $N_0 = N_1 + 1$ or $N_0 = N_1$ must be satisfied. One of the main conclusions of this section is that $N_1 = N_0$ is the essential truncation condition, at least for the expansion of the Poincar\'e or the space-time decomposed Poincar\'e superalgebra, to establish an off-shell supersymmetric representation for a larger superalgebra. The following sections, Section \ref{Section3} and Section \ref{Section4} mainly serve to provide expamples for the procedure that we describe in Section \ref{Section2}. In Section \ref{Section3}, we employ the procedure that we described in Section \ref{Section2} and find the (anti-)chiral multiplet of four-dimensional $\cN=1$ coadjoint Poincar\'e superalgebra.  In Section \ref{Section4}, we discuss the scalar multiplet of three-dimensional $\cN=2$ non-relativistic superalgebra for ${g} = (1,0)$ (which corresponds to the $\cN=2$ extended Bargmann superalgebra) and ${g} = (1,1)$ superalgebra. We provide our comments and conclusions in Section \ref{Comments}. 
		
		\section{The Lie Algebra Expansions of Matter Multiplets}\label{Section2}
		\paragraph{}	
		The Lie algebra expansion is a procedure that takes a small Lie algebra and produces larger ones by expanding the gauge fields that correspond to the generators of the small algebra.	However, in general, we want to consider a theory that contains both the gauge fields $A_\mu^I$ and the \textit{matter fields} $\Phi$. The purpose of this section is to describe the procedure that is necessary for the consistent expansion of the matter fields, especially when they represent the \textit{matter multiplets} of a superalgebra. To provide a self-contained treatment, let us briefly remind the reader about the basics of covariant quantities such as covariant derivatives and curvatures, which are the fundamental building blocks of the expansion of matter fields and multiplets. We refer \cite{Freedman:2012zz} for readers interested in a more detailed discussion on general gauge theories and covariant quantities.
		
		Consider an algebra $\mathfrak{g}$ with generators $T_I$. A matter field $\Phi$ that transforms non-trivially under the action of these generators have the following transformation rules
		\begin{eqnarray}
			\delta(\epsilon) \Phi &=& [\epsilon^I T_I, \Phi] \,,
		\end{eqnarray}
		where $\epsilon^I$ represents the transformation parameters corresponding to generators $T_I$. With this definition, the closure of the algebra on the field $\Phi$ is satisfied, that is, the commutator of any two sequential action on the field, $[\delta(\e_1),\delta(\e_2)]$ generates a third transformation, by use of the Jacobi identity
		\begin{eqnarray}\label{Jacobi}
			[\e^I_1 T_I, [ \e^J_2 T_J,\Phi]] - [ \e^J_2 T_J, [\e^I_1 T_I,\Phi]]  &=& [[\e^I_1 T_I, \e^J_2 T_J],\Phi] \,,
		\end{eqnarray} 
		where the commutator $[T_I, T_J]$ generates the necessary structure constants that relates the parameters $\e_1$ and $\e_2$ to a third parameter $\e_3^K = f_{IJ}{}^K \e_{1}^I \e_2^J$. Note that here, it is assumed that the closure of the algebra on $\Phi$ is \textit{off-shell} in the sense that no equation of motion have been used.	Next, we associate a gauge field to each of the generators, i.e. $A_\mu = T_I A_\mu^I$. Then, the covariant derivative of the matter field $\Phi$ is given by
		\begin{eqnarray}\label{CovDer1}
			\mathcal{D}_\mu \Phi = \partial_\mu \Phi - \delta(A_\mu) \Phi \,.
		\end{eqnarray}
		Here $ \delta (A_\mu^I)$ means that one needs to calculate the transformation of the field $\Phi$ with respect to all gauge symmetries generated by $T_I$ and replace the relevant transformation parameters with the gauge fields $A_\mu^I$. Consequently, the commutator of two covariant derivatives on the matter field $\Phi$ takes the following form
		\begin{eqnarray}\label{DD1}
			[\mathcal D_\mu, \mathcal D_\nu] \Phi = - \delta(R_{\mu\nu}) \Phi \,,
		\end{eqnarray}	
		where the group-theoretical curvature $R_{\mu\nu}{}^I$ is defined as
		\begin{eqnarray}
			R_{\m\n}{}^I &=& 2 \partial_{[\mu}  A_{\nu]}^I + f_{JK}{}^I A_\mu^K A_\mu^J \,.
		\end{eqnarray}
		These definitions of covariant derivative \eqref{CovDer1} and commutators \eqref{DD1} are problematic if the algebra $\mathfrak{g}$ is associated with spacetime symmetries. For instance, with the definition of the covariant derivatives \eqref{CovDer1}, the covariant derivative of a scalar field simply vanishes for a gauge group that includes local translations and if the scalar field carries no internal symmetry. To evade this problem, the definition of covariant derivative  \eqref{CovDer1} is understood as $ \delta (A_\mu^I)$ includes all gauge transformations except $P-$gauge transformations. The compensation for the exclusion of the $P-$gauge transformations for the commutation relations, \eqref{DD1}, is that we need to impose the vanishing of the curvature of local translations, i.e., $R_{\mu\nu}{}^a (P) = 0$. However, in an expansion procedure, setting $R_{\mu\nu}{}^a (P) = 0$ before expanding the gauge fields and the matter fields would lead us to miss any contribution to the transformation rules that arise from the expansion of the generators of local translations. Thus, we modify the structure of \eqref{DD1} as
		\begin{eqnarray}\label{DD2}
			[\mathcal D_\mu, \mathcal D_\nu] \Phi = - \delta(R_{\mu\nu}) \Phi |_{R(P)=0} \,.
		\end{eqnarray}	
		which means that we first perform any necessary calculation on the right-hand side of this equation and impose the constraint $R_{\mu\nu}{}^a (P) = 0$ only at the final stage.
		
		\subsection{The General Procedure}
		\paragraph{}
		Let us now consider that the algebra $\mathfrak{g}$ can be split into the direct sum $\mathfrak{g} = V_0 \oplus V_1$ where $V_0$ represent the subspace of even generators while $V_1$ represent the subspace of odd generators. In this case, the generators of this algebra $T_I$ can be split as $T_I = \{X_i, Y_\alpha\}$ with $X_i \in V_0$ and $Y_\alpha \in V_1$.  The transformation rule for a matter field $\Phi$ is then given by
		\begin{eqnarray}\label{Transform}
			\delta \Phi &=& [\e^i X_i, \Phi] + [\e^\alpha Y_\alpha, \Phi]  \,,
		\end{eqnarray}
		where the commutation relations for the generators $T_I = \{X_i, Y_\alpha\}$ takes the following form
		\begin{align}
			[X_i, X_j] & = f_{ij}{}^j X_k \,, & [Y_\alpha, Y_{\beta}] & = f_{\alpha\beta}{}^i X_i \,, & [X_i, Y_\alpha] & = f_{i\alpha}{}^\beta Y_\beta \,,
		\end{align}
		due to the splitting of the algebra into even and odd parts with respect to \eqref{EvenOdd}. In this case, the Jacobi identity that is necessary for the off-shell closure of the algebra on the field $\Phi$ decomposes as
		\begin{align}\label{ExpandedJacobi}
			[\e_1^i X_i, [\e_2^j X_j, \Phi]] - 	[\e_2^j X_j, [\e_1^i X_i, \Phi]] & = [[\e_1^i X_i, \e_2^j X_j], \Phi] \,,\nonumber\\
			[\e_1^\a Y_\a, [\e_2^\b Y_\b, \Phi]] - 	[\e_2^\b Y_\b, [\e_1^\a Y_\a, \Phi]] & = [[\e_1^\a Y_\a, \e_2^\b Y_\b], \Phi] \,,\nonumber\\
			[\e^i X_i, [\e^\a Y_\a, \Phi]] - 	[\e^\a Y_\a, [\e^i X_i, \Phi]] & = [[\e^i X_i, \e^\a Y_\a], \Phi] \,.
		\end{align}
		Similarly, the commutator of two covariant derivatives are split into even and odd part as
		\begin{eqnarray}\label{DD3}
			[\mathcal D_\mu , \mathcal D_\nu] \Phi = \left( - \d(R_{\mu\nu}{}^i) \Phi -  \d(R_{\mu\nu}{}^\alpha) \Phi \right)|_{R(P)=0} \,.
		\end{eqnarray}
		The form of transformation rules \eqref{Transform}, the Jacobi identities \eqref{ExpandedJacobi} and the commutation relations \eqref{DD3} are our starting point to determine the expansion character, therefore the expansion of the matter field $\Phi$. We may now consider the expansion of the gauge fields according to \eqref{ExpansionGaugeFields}, which gives rise to the following set of commutation relations among the generators of the larger algebra
		\begin{align}\label{ExpandedAlgebra}
			[X_i^{(2m)}, X_j^{(2n)}] & = f_{ij}{}^j X_k^{(2m+2n)} \,, & [Y_\alpha^{(2m+1)}, Y_{\beta}^{(2n+1)}] & = f_{\alpha\beta}{}^i X_i^{(2m+2n+2)} \,, \nonumber\\
			[X_i^{(2m)}, Y^{(2n+1)}_\alpha] & = f_{i\alpha}{}^\beta Y_\beta^{(2m+2n+1)} \,,
		\end{align}
		where we defined the expanded gauge field as
		\begin{eqnarray}
			A = A^i_{(2n)} X_i^{(2n)} + A^\alpha_{(2n+1)} Y_\alpha^{(2n+1)} \,.
		\end{eqnarray}
		With this expansion of the gauge fields, curvatures and algebra, let us work out the details of the expansion of the gauge fields and matter fields by providing examples for relativistic and non-relativistic cases.
		
		\subsubsection{Coadjoint Construction}\label{CoAd}
		\paragraph{}
		In \cite{Bergshoeff:2020fiz}, it was shown that three-dimensional non-relativistic gravity models may be obtained from the contractions of Poincar\'e, Poincar\'e $\oplus\,\mathfrak{u}(1)^2$, coadjoint Poincar\'e or coadjoint Poincar\'e $\oplus\, \mathfrak{u}(1)^2$ algebra. The models that go beyond the Galilei algebra are associated with the coadjoint Poincar\'e algebra which may be obtained by a certain Lie algebra expansion of the Poincar\'e algebra itself.
		
		The $D$-dimensional Poincar\'e algebra has the following form for its non-vanishing commutators
		\begin{eqnarray}
			[P, J] \sim P \,, \qquad [M, M] \sim M \,,
		\end{eqnarray}
		where $P_a$ represents the generator of translations while $M_{ab}$ is the generator of Lorentz transformations. This structure of the algebra allows us to classify the generators in two ways
		\begin{enumerate}
			\item{Both $P_a$ and $M_{ab}$ are in even class, i.e. $V_0 = \{P_a, M_{ab}\}$. As the Lie algebra expansion preserves the structure constants \eqref{ExpandedAlgebra}, we may give the form of the expanded algebra as
				\begin{eqnarray}
					[P, J] \sim P \,, \qquad [M, M] \sim M \,, \qquad  [T,J] \sim T\,, \qquad [P, S] \sim T \,, \qquad [J,S] \sim S \,,
				\end{eqnarray}
				where we made the following identifications
				\begin{eqnarray}
					P_a^{(0)} = P_a\,, \qquad P_a^{(2)} = T_a \,, \qquad J_{ab}^{(0)} = J_{ab} \,, \qquad J_{ab}^{(2)} = S_{ab} \,.
				\end{eqnarray}
				This is the coadjoint Poincar\'e algebra. This form of expansion has been useful in finding non-relativistic algebras. When the Poincar\'e algebra is improved with a cosmological constant, i.e. $[P,P] \sim \Lambda M$ with $\L$ being the cosmological constant, such an expansion yields two additional non-vanishing commutation relations
				\begin{eqnarray}
					[P,P] \sim \L M \,, \qquad [P, T] \sim \L S \,.
				\end{eqnarray}
				which gives rise to the coadjoint AdS algebra. This form of the algebra has been useful in relating Newton-Hooke type algebras to coadjoint AdS algebra. It has also been an essential element to relate three-dimensional massive gravity models to an infinite-dimensional algebra \cite{Bergshoeff:2021tbz}. When this expansion is generalized to include supersymmetry, the $D$-dimensional Poincar\'e superalgebra brings the following two commutators for the SUSY generators
				\begin{eqnarray}\label{RelativisticSUSY}
					\{Q,Q\} \sim P \,, \qquad [M,Q] \sim Q \,,
				\end{eqnarray}
				where we ignore possible central charges. In this case, there is no restriction on $Q$ as the multiplication table \eqref{EvenOdd} allows for both even and odd choice of $Q$. Thus, $Q$ may be expanded in all powers of $\lambda$ and the consistent truncation yields an extended superalgebra.
			}
			\item{We may also split the generators of the Poincar\'e algebra as $V_0 = \{M_{ab}\}$ and $V_1 = \{P_a\}$. In this case, the first ${g} = (0,0)$ order algebra is the Poincar\'e algebra itself while ${g} = (1,1)$ algebra is the coadjoint Poincar\'e algebra
				\begin{eqnarray}
					[P, J] \sim P \,, \qquad [M, M] \sim M \,, \qquad  [T,J] \sim T\,, \qquad [P, S] \sim T \,, \qquad [J,S] \sim S \,,
				\end{eqnarray}
				Note that in this case, the cosmological commutators are different, 
				\begin{eqnarray}
					[P,P] \sim \L  S \,.
				\end{eqnarray}
				So far, there is no known relation for this form of the expansion to a Newton-Hooke type of a non-relativsitic algebra. The form of supersymmetry commutators \eqref{RelativisticSUSY} is more subtle then the even/even expansion. As $P$ belongs to odd class of generators, the $\{Q,Q\} \sim P$ anti-commutator implies that the supersymmetry generator must first be projected into at least two independent parts, say $Q_\pm$, in a way that only the opposite projections have a non-vanishing commutator $\{Q_+, Q_-\} \sim P$. We may then assign opposite expansion characters to $Q_+$ and $Q_-$ to preserve the structure of supersymmetry, i.e. two sequential SUSY transformations generate translations.
			}
		\end{enumerate}
		Ignoring the cosmological constant, the relation between the coadjoint Poincar\'e algebra and the three-dimensional non-relativistic algebras suggest that it may hold in supersymmetric settings, in various dimensions, and with higher-order expansions.	With these two different ways to generate larger algebras from super-Poincar\'e algebra, let us briefly discuss the character of matter multiplets. 
		\begin{enumerate}
			\item{Let's first assume that  $V_0 = \{P_a, M_{ab}\}$ in which case the SUSY generator could belong to both $V_0$ or $V_1$. For instance, for four-dimensional $\cN = 1$ supersymmetry, we have
				\begin{eqnarray}
					\{P_L Q , \overline{P_R Q}\} \sim (P_L \gamma^a) P_a \,,
				\end{eqnarray}  
				thus, both $\{P_L Q, P_R Q\} \in V_0$ or $\{P_L Q, P_R Q\} \in V_1$ are valid choices. However, choosing both to be in $V_1$ means that at the lowest order, two sequential SUSY transformations do not generate translations. Thus, we choose both generators to be an element of $V_0$. Consequently, all matter fields can be expanded in even powers of an expansion parameter $\lambda$. As all generators, including the SUSY generators, as well as the matter fields are expanded in even powers, this turns into an identical procedure presented for $1/c$ expansion in \cite{Hansen:2020pqs}.
			}
			\item{Next, let's assume that $V_0 = \{P_a\}$ and $V_1 = \{M_{ab}\}$. As mentioned, this would require one to project the SUSY generator $Q$ into two independent parts and assign opposite character to different projections to preserve $\{Q,Q\} \sim P$ structure. Following the four-dimensional $\cN = 1$ expample, we need to assign opposite expansion character to $P_L Q$ and $P_R Q$, e.g. $P_L Q \in V_0$ while $P_R Q \in V_1$. This splitting implies that we must set $N_0 = N_1$ in order to have as many left-projected generators as right-projected ones. As we will discuss in Section \ref{Section3}, setting $N_1 = N_0 + 1$ also fails the expansion of the Jacobi identities \eqref{ExpandedJacobi}. To see the behavior of the fields, let us consider a complex scalar $Z$ field that transforms to a left-projected spinor $P_L \chi$, 
				\begin{eqnarray}
					\d Z = \bar\e P_L \chi \,,
				\end{eqnarray}   
				Then, the contribution of supersymmetry transformation to the commutator of covariant derivatives is
				\begin{eqnarray}
					[\cD_\m, \cD_\n] Z \sim \bar{R}_{\m\n} (Q) P_L \chi \,.
				\end{eqnarray}
				Since we choose $P_L {R}_{\m\n}(Q)$ to be expanded in even powers, then $Z$ and $P_L \chi$ can both be even or odd. The expansion character for the remaining fields of the multiplet can be investigated by applying the same idea to higher order components. For an anti-chiral multiplet, however, have a complex scalar $\bar Z$ that transforms to a right-projected spinor
				\begin{eqnarray}
					\d \bar{Z} = \bar\e P_R \chi \,,
				\end{eqnarray}
				which implies
				\begin{eqnarray}
					[\cD_\m, \cD_\n] Z \sim \bar{R}_{\m\n} (Q) P_R \chi \,.
				\end{eqnarray}
				For this case, as choose $P_R {R}_{\m\n}(Q)$ to be expanded in odd powers, we must chose opposite expansion characters for $\bar{Z}$ and $P_R \chi$. Once again, the remaining fields of the multiplet can be investigated by performing a similar analysis for the higher order components.}
		\end{enumerate}
		
		\subsubsection{Spacetime Decomposition}
		\paragraph{}
		In \cite{Bergshoeff:2019ctr}, it was shown that if the generators of the Poincar\'e algebra is split into space and time components, then the resulting decomposed algebra can be written as a direct sum of $V_0$ and $V_1$ where
		\begin{eqnarray}
			V_0 = \{P_0, M_{ab}\}\,, \qquad V_1 = \{P_a, M_{a0}\} \,,
		\end{eqnarray}
		where we split the $D$-dimensional spacetime index $A$ as $A = (0,a)$ where $a$ represents the $(D-1)$-dimensional spatial index. The Lie algebra expansion of this spacetime decomposed algebra in three and four-dimensions then generate various known non-relativistic algebras. In order to reflect such a decomposition to supersymmetry generators, one first need to find projection operator(s) such that the projected SUSY generators give rise to both $P_a$ and $H$. Otherwise, we do not have a superalgebra in the sense that the anti-commutator of two SUSY generators do not lead to diffeomorphisms. Using the same projection operators, one may then implement a spacetime splitting on any off-shell multiplet of the Poincar\'e superalgebra. As the bosonic fields are split into both even and odd classes, the expansion of the spacetime decomposed off-shell multiplet follows the same pattern that we discussed for the coadjoint construction with even/odd splitting.

		
		\section{Four-Dimensional $\cN=1$ Co-adjoint Poincar\'e Superalgebra} \label{Section3}
		\paragraph{}
		In this section, we provide the expansion of the four-dimensional $\cN=1$ off-shell (anti-)chiral multiplet an example for the coadjoint construction. The four-dimensional super-Poincar\'e algebra is given by
		\begin{align}
			[P_{A}\,, M_{BC}]&= \eta_{A B} P_{C}- \eta_{A C} P_{B}\,, &		\{P_LQ ,\overline{P_RQ} \}&= - \frac{1}{2} P_L\Gamma^{A}\ P_{A}  \,,\nn\\		
			[M_{A B},P_LQ]&=-\frac{1}{2} \Gamma_{A B}\ P_LQ \,,&	[M_{A B},P_RQ]&=-\frac{1}{2} \Gamma_{A B}\ P_RQ  \,, 
		\end{align}
		as well as
		\begin{align}
			[M_{A B},M_{C D }]&= \eta_{A C } M_{D B} - \eta_{B C } M_{D A} -\eta_{A D } M_{C B} + \eta_{B D} M_{C A}  \,.	
		\end{align}
		Here, $A$ is the spacetime index $A= 0,1,2,3$ and we use $\eta_{A B} = \text{diag}(-,+,+,+)$. Furthermore, we define
		\begin{eqnarray}
			P_L = \frac12 \left( \mathbb{I} + \Gamma_\star \right) \,, \qquad P_R = \frac12 \left( \mathbb{I} - \Gamma_\star \right) \,, 
		\end{eqnarray}
		where $\Gamma_\star = \rmi \Gamma_0 \Gamma_1 \Gamma_2 \Gamma_3$. As mentioned, we can assign all the generators an even expansion character which is consistent, but it is the same as $1/c$-expansion of \cite{Hansen:2020pqs}, thus we skip that possibility here. However, we may also set
		\begin{eqnarray}\label{SplitSuperPoincare}
			V_0 = \{M_{AB}, P_L Q\}\,, \qquad V_1 = \{P_A, P_R Q\} \,.
		\end{eqnarray}
		Following the standard Lie algebra expansion procedure, we may use \eqref{ExpandedAlgebra}, which applies to both commutators and anti-commutators, and obtain the following non-vanishing (anti-)commutators for the four-dimensional $\cN=1$ coadjoint Poincar\'e superalgebra
		\begin{align}
			[P_{A}\,, M_{BC}]&= \eta_{A B} P_{C}- \eta_{A C} P_{B}\,, &	[T_{A}\,, M_{BC}]&= \eta_{A B} T_{C}- \eta_{A C} T_{B}\nn \\
			[P_{A}\,, S_{BC}]&= \eta_{A B} T_{C}- \eta_{A C} T_{B} \,,  & [M_{A B},P_L Q]&=-\frac{1}{2} \Gamma_{A B}\ P_L Q \,, \nonumber\\
			[S_{A B},P_L Q]&=-\frac{1}{2} \Gamma_{A B}\ P_L S \,, & [M_{A B},P_L S]&=-\frac{1}{2} \Gamma_{A B} P_L S\,,\nonumber\\
			[M_{A B},P_R Q]&=-\frac{1}{2} \Gamma_{A B}\ P_R Q \,, & [M_{A B},P_R S]&=-\frac{1}{2} \Gamma_{A B}\ P_R S  \,,\nonumber\\ 
			[S_{A B},P_R Q]&=-\frac{1}{2} \Gamma_{A B} P_R S \,, &	\{P_LQ ,\overline{P_RQ} \}&= - \frac{1}{2} P_L\Gamma^{A}\ P_{A} \,,\nonumber\\
			\{P_LQ ,\overline{P_R S} \}&= - \frac{1}{2} P_L\Gamma^{A}\ T_{A} \,, & 	\{P_L S ,\overline{P_RQ} \}&= - \frac{1}{2} P_L\Gamma^{A}\ T_{A} \,,
		\end{align}
		as well as
		\begin{align}
			[M_{A B},M_{C D }]&= \eta_{A C } M_{D B} - \eta_{B C } M_{D A} -\eta_{A D } M_{C B} + \eta_{B D} M_{C A}\,, \nn \\ 
			[S_{A B},M_{C D }]&= \eta_{A C } S_{D B} - \eta_{B C } S_{D A} -\eta_{A D } S_{C B} + \eta_{B D} S_{C A} \,.
		\end{align}
		Note that here we set
		\begin{align}\label{CoAdNewGen}
			P_A^{(1)} &= P_A \,, & P_A^{(3)} &= T_A \,, & M_{AB}^{(0)} &= M_{AB} \,, & M_{AB}^{(2)} &= S_{AB} \,, \nonumber\\
			P_L Q^{(0)} & = P_L Q \,, & 	P_L Q^{(2)} & = P_L S \,,  & 	P_R Q^{(1)} & = P_R Q \,, & 	P_R Q^{(3)} & = P_R S \,. 
		\end{align}	
		With these expressions for the generators in hand, it is possible establish a relation between the three-dimensional $\cN=2,\, g=(1,1)$ superalgebra and the $\cN=1$ coadjoint Poincar\'e superalgebra by means of contraction. For the bosonic part, the necessary contraction has already been found in \cite{Bergshoeff:2020fiz}. This can, in principle, be generalized to include the $[B,F]$ and $\{F,F\}$ (anti-)commutators of the superalgebra. Once this is achieved, one can take the contraction of the off-shell multiplets of $\cN=1$ coadjoint Poincar\'e superalgebra to produce the multiplets of $\cN=2,\, g=(1,1)$ superalgebra. In particular, noticing that the contraction and the Lie algebra expansion coincides at the lowest order, one can simply take the multiplets of $\cN=1$ space-time decomposed coadjoint Poincar\'e superalgebra and consider its lowest order expansion with respect to the classification of the generators to match with the three-dimensional $\cN=2,\, g=(1,1)$ superalgebra. Thus, the results of this section can be regarded as a necessary preliminary step to establish the non-relativistic matter multiplets and actions. With this note in mind, we now proceed to the expansion of the off-shell (anti-)chiral multiplets of four-dimensional $\cN=1$ Poincar\'e superalgebra.
		
		\subsection{(Anti-)Chiral Multiplet}
		\paragraph{}
		The four-dimensional off-shell rigid $\cN=1$ chiral multiplet has the following transformation rules
		\begin{eqnarray} \label{Chiral}
			\d   Z &=& \x^A \partial_A Z + \frac{1}{\sqrt 2} \bar \epsilon P_L \chi \nonumber \\
			\d P_L \chi &=& \x^A \partial_A P_L \chi + \frac12 \l^{AB} \G_{AB} P_L \chi + \frac{1}{\sqrt 2} \G^A \partial_A   Z  P_R \epsilon +   \frac{1}{\sqrt 2} F P_L \epsilon \nonumber\\
			\d   F &=& \x^A \partial_A F +\frac{1}{\sqrt 2} \bar \epsilon \, \G^A \partial_A P_L \chi \,,  
		\end{eqnarray}
		Here, $\x^A$ is the parameter for translations, $\l^{AB}$ for Lorentz transformations and $P_L \e$ (or $P_L \e$) for left-projected (or right-projected) supersymmetry transformations. The transformation rules for the anti-chiral multiplet, which is the complex conjugate of the chiral multiplet, is given by
		\begin{eqnarray}\label{AntiChiral}
			\d \bar Z &=& \x^A \partial_A \bar Z + \frac{1}{\sqrt 2} \bar \epsilon P_R \chi \nonumber \\
			\d P_R \chi &=&\x^A \partial_A P_R \chi + \frac12 \l^{AB} \G_{AB} P_R \chi + \frac{1}{\sqrt 2}  \G^A \partial_A \bar Z  P_L \epsilon + \frac{1}{\sqrt 2} \bar F P_R \epsilon \nonumber \\
			\d \bar F &=& \x^A \partial_A \bar F + \frac{1}{\sqrt 2} \bar \epsilon \, \G^A \partial_A P_R \chi \,.  
		\end{eqnarray}
		Following the steps that we present in Section \ref{CoAd}, we need to establish the commutator of covariant derivatives and use vielbein, or its inverse, to convert the flat indices on the derivatives to curved indices. We can then use the leading order scaling or the vielbein, which, according to the splitting \eqref{SplitSuperPoincare} is given by 
		\begin{eqnarray}
			E_\m{}^A  = \lambda e_\m{}^A \,, \qquad E^\m{}_A = \frac1{\lambda} e^\m{}_A \,,
		\end{eqnarray}
		and establish the expansion character of the fields of the (anti-)chiral multiplet. Here, $e_\m{}^A$ is the gauge field associated with the leading order term in the expansion of $P_A$, which we chose to be the generator of translation in the coadjoint Poincar\'e superalgebra. Note that once the expansion is complete, we set $e_\m{}^A = \d_\m{}^A$ and obtain the rigid chiral multiplet of four-dimensional $\cN=1$ coadjoint Poincar\'e algebra on flat background.
		
		\subsubsection{Expansion of  Chiral Multiplet}
		\paragraph{}
		Let us start with the  transformations of the chiral multiplet \eqref{Chiral} to see how the new generators, $T_A, S_{AB}, P_L S$ and $P_R S$ act on the components of a chiral superfield. Converting the flat indices on derivatives to curved indices, the transformation rules for the chiral multiplet read
		\begin{eqnarray} \label{ChiralBose}
			\d   Z &=& \x^A E^\mu{}_A \partial_\mu Z  + \frac{1}{\sqrt 2} \bar \epsilon P_L \chi \nonumber \\
			\d P_L \chi &=& \x^A  E^\mu{}_A \partial_\mu P_L \chi + \frac12 \l^{AB} \G_{AB} P_L \chi + \frac{1}{\sqrt 2} \G^A \partial_A   Z  P_R \epsilon +   \frac{1}{\sqrt 2} F P_L \epsilon \nonumber\\
			\d   F &=& \x^A  E^\mu{}_A \partial_\mu F+\frac{1}{\sqrt 2} \bar \epsilon P_R \, \G^A E^\m{}_A \partial_\m P_L \chi \,,  
		\end{eqnarray}
		Note that we did not turn the partial-derivative into covariant derivative since in the rigid limit, all gauge fields except $E_\m{}^A$ and its inverse are set to zero. In order to expand the fields, we need to know their expansion character. This can be understood by the structure of the chiral superfield $\Phi$
		\begin{eqnarray}
			\Phi = Z + \frac1{\sqrt2} \overline{P_L \theta} P_L \chi + \frac14 \overline{P_L \theta} P_L \theta F \,,
		\end{eqnarray}
		where $\theta$ represents the anti-commuting coordinates. Based on the character of the associated translation generator, $P_L Q$, we may expand $P_L \e$ as
		\begin{eqnarray}
			P_L \e \quad \to \quad  P_L \e + \l^2 P_L \eta \,,
		\end{eqnarray}
		where $P_L \eta$ is associated with the new left-projected generator, $P_L S$ and represents additional anti-commuting coordinates due to the increasing number of supercharges. As $g=(0,0)$ truncation is the Poincar\'e superalgebra itself, we do not wish to spoil the zeroth-order definition of the superfield $\Phi$ so that the truncation of extra generators consistently yield a chiral multiplet. Thus, we expand the fields as
		\begin{align}\label{ExpansionChiral}
			Z = & Z_1 + \l^2 Z_2 \,, & P_L \chi &=  P_L \chi_1 + \l^2 P_L \chi_2 \,, & F & = F_1 + \l^2 F_2 \,,
		\end{align} 
		which would lead us to the expansion of the superfield, $\Phi = \Phi_0 + \l^2 \Phi_2$ where
		\begin{eqnarray}\label{ExpChiralSuperfield}
			\Phi_0 &=& Z_1 + \frac1{\sqrt2} \overline{P_L \theta} P_L \chi_1 + \frac14 \overline{P_L \theta} P_L \theta F_1 \,,\nonumber\\
			\Phi_2 &=& Z_2 + \frac1{\sqrt2} \overline{P_L \zeta} P_L \chi_1 + \frac1{\sqrt2} \overline{P_L \theta} P_L \chi_2 + \frac12 \overline{P_L \theta} P_L \zeta F_1 + \frac14 \overline{P_L \theta} P_L \theta F_2 \,.
		\end{eqnarray}
		Here, $P_L \zeta$ is the additional anti-commuting coordinates that arise from the expansion of $P_L \theta$, i.e. $P_L \theta \to P_L \theta + \l^2 P_L \zeta$. Note that $\Phi_2$ itself is not a superfield. As we will discuss momentarily, we may think of $\Phi_0$ and $\Phi_2$ as entries of a matrix superfield. The new generators, $T_A, S_{AB}, P_L S$ and $P_R S$ act on these superfields in a way that while they have the same operator structure as the generators that they originate from, they also have a matrix character such that they act on the matrix superfield and transform the entries of $\Phi_2$ into the entries of $\Phi_0$. To see this in action, let us  consider the transformation rules for the complex scalar field $Z$, which gives rise to the following the commutator of covariant derivatives
		\begin{eqnarray}
			[\cD_\m, \cD_\n] Z &=& \left( - R_{\m\n}{}^A (P) E_A{}^\m \partial_\m Z - \overline{P_L {R}}_{\m\n}(Q) P_L \chi \right)|_{R(P)=0}  \,,
		\end{eqnarray}
		where we used the four-dimensional identity $\bar{\l} P_L = \overline{P_L \l}$. Expanding the fields as in \eqref{ExpansionChiral}, using the zeroth-order scaling character of the vielbein, $E_\m{}^A = \lambda e_\m{}^a$ and $E^\m{}_A = \frac{1}{\lambda}\ e^\m{}_a$ 	and expand the curvatures with respect to their expansion characters \eqref{SplitSuperPoincare}, we obtain
		\begin{align}
			[\cD_\m, \cD_\n] Z_1 &=  \left( - R_{\m\n}{}^A (P) E_A{}^\m \partial_\m Z_1 - \frac{1}{\sqrt 2}  \overline{P_L {R}}_{\m\n}(Q) P_L \chi_1 \right)|_{R(P)=0} \,,\nonumber\\
			[\cD_\m, \cD_\n] Z_2 &=  \Big( - R_{\m\n}{}^A (P) E_A{}^\m \partial_\m Z_2  - R_{\m\n}{}^A (T) E_A{}^\m \partial_\m Z_1  - \frac{1}{\sqrt 2}  \overline{P_L {R}}_{\m\n}(Q) P_L \chi_2  \nonumber\\
			& \qquad -\frac{1}{\sqrt 2}   \overline{P_L {R}}_{\m\n}(S) P_L \chi_1 \Big)|_{R(P)=0} \,.
		\end{align}
		Imposing the constraint $R_{\m\n}{}^a (P) = 0$ and setting $E^\m{}_A = \d^\m{}_A$, we obtain the transformation rules for $Z_1$ and $Z_2$ as
		\begin{eqnarray}
			\d Z_1 &=& \x^A  \partial_A Z_1 +\frac{1}{\sqrt 2}  \bar\e P_L \chi_1 \,,\nonumber\\
			\d Z_2 &=& \x^A  \partial_A Z_2 + \l_T^A \partial_A Z_1  + \frac{1}{\sqrt 2}  \bar\e P_L \chi + \frac{1}{\sqrt 2}  \bar\eta P_L \chi_2 \,.
		\end{eqnarray}
		Here, $\l_T^A$ are the transformation parameter for $T_A$-transformations while $P_L \eta$ is the parameter for $P_L S$-transformations. As evident from these transformation rules, $T_A$ has the same operator character as $P_A$ but it is a matrix of the operator $\partial_A$ since it turns the second order complex scalar $Z_2$ to the zeroth-order $Z_1$. We may, therefore, give the new supersymmetry generator $S_\alpha$ as a differential operator as
		\begin{eqnarray}
			S_\alpha  &=& S_\alpha^0 - \frac14 \left(\Gamma^A \theta \right)_\alpha T_A - \frac14 \left(\Gamma^A \zeta \right)_\alpha P_A \,,
		\end{eqnarray} 
		where $S_\alpha^{0}$ represents the generators for the shift $\zeta \to \zeta - c$, i.e. 
		\begin{eqnarray}
			S_\alpha^0 =  \frac{\partial}{\partial \bar\zeta}
		\end{eqnarray}
		Next, we turn the transformation rules for $P_L \chi$, which is given by
		\begin{eqnarray}
			\d P_L \chi &=& \x^A E^\m{}_A \partial_\m P_L \chi + \frac12 \l^{AB} \G_{AB} P_L \chi + \frac{1}{\sqrt 2}  \G^A E^\m{}_A \partial_\m   Z  P_R \epsilon +  \frac{1}{\sqrt 2}  F P_L \epsilon\,.
		\end{eqnarray}
		which gives rise to 
		\begin{eqnarray}
			[\cD_\m, \cD_\n] P_L \chi &=& \Big( - R_{\m\n}{}^A(P) E^\m{}_A \partial_\m P_L \chi - \frac12 R_{\m\n}{}^{AB}(M) \G_{AB} P_L \chi \nonumber\\
			&& \quad - \frac{1}{\sqrt 2} \G^A E^\m{}_A \partial_\m Z P_R R_{\m\n}(Q) - \frac{1}{\sqrt 2} F P_L R_{\m\n}(Q)  \Big)|_{R(P) = 0}
		\end{eqnarray}
		Once again, expanding the fields as in \eqref{ExpansionChiral} and the curvatures with respect to their expansion characters \eqref{SplitSuperPoincare}, we obtain
		\begin{eqnarray}
			[\cD_\m, \cD_\n] P_L \chi_1 &=& \Big( - R_{\m\n}{}^A(P) E^\m{}_A \partial_\m P_L \chi_1 - \frac12 R_{\m\n}{}^{AB}(M) \G_{AB} P_L \chi_1   \nonumber\\
			&& \quad -\frac{1}{\sqrt 2}  \G^A E^\m{}_A \partial_\m Z_1 P_R R_{\m\n}(Q) - \frac{1}{\sqrt 2} F_1 P_L R_{\m\n}(Q)  \Big)|_{R(P) = 0}
		\end{eqnarray}
		We may follow the same procedure that we implemented for the expansion of $Z$ and use the zeroth-order scaling character of the vielbein as well as the expansion of the fields \eqref{ExpansionChiral} and the curvatures with respect to \eqref{SplitSuperPoincare}. This gives rise to the following transformation rules for the left-projected spinors $P_L \chi_1$ and $P_L \chi_2$
		\begin{eqnarray}
			\delta P_L \chi_1 &=& \xi^A \partial_A  P_L \psi_1 +\frac12 \l^{AB} \g_{AB} P_L \chi_1 +  \frac{1}{\sqrt 2}  \g^A \partial_A Z_1 P_R \e + \frac{1}{\sqrt 2} F_1 P_L \e  \,,\nonumber\\
			\delta P_L \chi_2 &=&  \xi^A \partial_A P_L \psi_2 +  \l^{A}_T \partial_{A} P_L  \psi_1+ \frac12 \l^{AB} \g_{AB} P_L \psi_2+ \frac12 \sigma^{AB} \g_{AB} P_L \psi_1  \nonumber\\
			&&+  \frac{1}{\sqrt 2} \g^A \partial_A Z_1 P_R \eta  +  \frac{1}{\sqrt 2} \g^A \partial_A Z_2 P_R \e + \frac{1}{\sqrt 2} F_2 P_L \e +  \frac{1}{\sqrt 2} F_1 P_L \eta \,.
		\end{eqnarray} 
		Note that once again, $T_A$ transforms $P_L \chi_2$ to $P_L \chi_1$ with the same differential operator form of $P_A$ Similarly, $S_{AB}$ acts on $P_L \chi_2$ in the same way as $M_{AB}$ but now with a matrix character that is $P_L \chi_2$ is transformed to $P_L \chi_1$. Finally, we may expand the auxiliary complex field $F$, which gives rise to the following transformation rules for the chiral multiplet of the $\cN=1$ coadjoint Poincar\'e algebra
		\begin{eqnarray}
			\d Z_1 &=& \x^A  \partial_A Z_1 +\frac{1}{\sqrt 2}  \bar\e P_L \chi_1 \,,\nonumber\\
			\d Z_2 &=& \x^A  \partial_A Z_2 + \l_T^A \partial_A Z_1  + \frac{1}{\sqrt 2}  \bar\e P_L \chi_2 + \frac{1}{\sqrt 2}  \bar\eta P_L \chi_1 \,, \nonumber\\
			\delta P_L \chi_1 &=& \xi^A \partial_A  P_L \chi_1 +\frac12 \l^{AB} \g_{AB} P_L \chi_1 +  \frac{1}{\sqrt 2}  \g^A \partial_A Z_1 P_R \e + \frac{1}{\sqrt 2} F_1 P_L \e  \,,\nonumber\\
			\delta P_L \chi_2 &=&  \xi^A \partial_A P_L \chi_2 +  \l^{A}_T \partial_{A} P_L  \chi_1+ \frac12 \l^{AB} \g_{AB} P_L \chi_2+ \frac12 \sigma^{AB} \g_{AB} P_L \chi_1  \nonumber\\
			&&+  \frac{1}{\sqrt 2} \G^A \partial_A Z_1 P_R \eta  +  \frac{1}{\sqrt 2} \G^A \partial_A Z_2 P_R \e + \frac{1}{\sqrt 2} F_2 P_L \e +  \frac{1}{\sqrt 2} F_1 P_L \eta \,,\nn\\
			\d F_1 &=& \x^A   \partial_A F_1 +\frac{1}{\sqrt 2} \bar \epsilon \, \G^A \partial_A P_L \chi_1   \,,\nn\\
			\d F_2 &=& \x^A   \partial_A F_2 + \l_T^A   \partial_A F_1  +\frac{1}{\sqrt 2} \bar \epsilon \, \G^A \partial_A P_L \chi_2 +  \frac{1}{\sqrt 2} \bar \eta \, \G^A \partial_A P_L \chi_1   \,.
		\end{eqnarray}
		With this result in hand, let us discuss some technical notes and noteworthy properties of this multiplet. First of all, we have explicitly checked by using \textit{GammaMaP} \cite{Kuusela:2019iok} that this multiplet is indeed closed off-shell given that the third parameters that are consistent with the $\cN=1$ coadjoint Poincar\'e algebra are given by
		\begin{align}
			\{Q_\a, Q_\b\} &= - \frac12( \Gamma^A)_{\a\b} P_A \,, & \Rightarrow & &  \x^A_3 &= - \frac12 \bar\epsilon_1 \Gamma^A \epsilon_2\nn\\
			\{Q_\a, {S}_\b\} &= - \frac12 ( \Gamma^A)_{\a\b}  T_A & \Rightarrow & &  \l^A_{T3} &= - \frac12 \bar\epsilon \, \Gamma^A \eta \,.
		\end{align}
		Second, this multiplet is not irreducible, that is, as evident from the expansion of the superfield \eqref{ExpChiralSuperfield}, the zeroth-order superfield can be truncated by setting $\{Z_1, P_L \chi_1, F_1\}$ to zero in which case $\Phi_2$ becomes identical to the chiral superfield. This is a typical property of the expansion: a lower-order field do not pick up extra transformation rules from higher-order generators in an expansion procedure \cite{Bergshoeff:2019ctr,Hansen:2020pqs}. For instance, the transformation rules of vielbein or spin-connection do not get modified under the expansion \eqref{CoAdNewGen}, see \cite{Bergshoeff:2019ctr} for non-relativistic examples. Finally, we may take the complex conjugate of this set of transformation rules, which gives rise to the following anti-chiral multiplet
		\begin{eqnarray}\label{ExpandedAntiChiral}
			\d \bar Z_1 &=& \x^A  \partial_A \bar Z_1 +\frac{1}{\sqrt 2}  \bar\e P_R \chi_1 \,,\nonumber\\
			\d \bar Z_2 &=& \x^A  \partial_A \bar Z_2 + \l_T^A \partial_A \bar Z_1  + \frac{1}{\sqrt2}  \bar\e P_R \chi_2 + \frac{1}{\sqrt 2}  \bar\eta P_R \chi_1 \,, \nonumber\\
			\delta P_R \chi_1 &=& \xi^A \partial_A  P_R \chi_1 +\frac12 \l^{AB} \g_{AB} P_R \chi_1 +  \frac{1}{\sqrt 2}  \g^A \partial_A \bar Z_1 P_R \e + \frac{1}{\sqrt 2} \bar F_1 P_R \e  \,,\nonumber\\
			\delta P_R \chi_2 &=&  \xi^A \partial_A P_R \chi_2 +  \l^{A}_T \partial_{A} P_R  \chi_1+ \frac12 \l^{AB} \g_{AB} P_R \chi_2+ \frac12 \sigma^{AB} \g_{AB} P_R \chi_1  \nonumber\\
			&&+  \frac{1}{\sqrt 2} \G^A \partial_A \bar Z_1 P_L \eta  +  \frac{1}{\sqrt 2} \G^A \partial_A \bar Z_2 P_L \e + \frac{1}{\sqrt 2} \bar F_2 P_R \e +  \frac{1}{\sqrt 2}\bar  F_1 P_R \eta \,,\nn\\
			\d\bar  F_1 &=& \x^A   \partial_A \bar F_1 +\frac{1}{\sqrt 2} \bar \epsilon \, \G^A \partial_A P_R \chi_1   \,,\nn\\
			\d \bar F_2 &=& \x^A   \partial_A \bar F_2 + \l_T^A   \partial_A \bar F_1  +\frac{1}{\sqrt 2} \bar \epsilon \, \G^A \partial_A P_R \chi_2 +  \frac{1}{\sqrt 2} \bar \eta \, \G^A \partial_A P_R \chi_1   \,.
		\end{eqnarray}
		
		\subsubsection{Expansion of Anti-Chiral Multiplet}
		\paragraph{}
		The anti-chiral multiplet for $\cN=1$ coadjoint Poincar\'e algebra can also be obtained by the direct expansion of the anti-chiral superfield $\bar \Phi$
		\begin{eqnarray}
			\bar\Phi = \bar Z + \frac{1}{\sqrt 2} \bar\theta P_R \chi + \frac14 \overline{P_R \theta} P_R \theta \bar F \,.
		\end{eqnarray}
		As $P_R \e$ is expanded in odd-powers of $\lambda$, i.e. $P_R \e  \to  \l P_R \e + \l^3 P_R \eta$,
		we need to choose the following expansion character for the components of the anti-chiral multiplet
		\begin{align}
			\bar Z & = \l \bar Z_1 + \l^3 \bar Z_2 \,, & P_R \chi & = P_R \chi_1 + \l^2 P_R \chi_2 \,, & \bar F = \frac{1}{\lambda} \bar F_1 + \l \bar F_2 \,.
		\end{align} 
		Here, it is important to note that $\bar F$ must start with the inverse power of $\l$, otherwise the lowest-order superfield is spoiled and one cannot realize the off-shell closure of the algebra on the expanded fields. This is an example of a mixed expansion in the sense that some of the fields are expanded in even powers while others are in odd powers in $\l$. With this choice of parametrization, the superfield $\bar\Phi$ is expanded as $\bar\Phi = \l \bar\Phi_0 + \l^3 \bar\Phi_1$ 
		where
		\begin{eqnarray}
			\bar\Phi_0 &=& \bar Z_1 + \frac{1}{\sqrt 2} \bar\theta P_R \chi_1 + \frac14 \overline{P_R \theta} P_R \theta \bar F_1 \,,\nn\\
			\bar\Phi_1 &=& \bar Z_2 + \frac{1}{\sqrt 2} \bar\theta P_R \chi_2 + \frac{1}{\sqrt 2} \bar\zeta P_R \chi_1  + \frac14 \overline{P_R \theta} P_R \zeta \bar F_1  + \frac14 \overline{P_R \theta} P_R \theta \bar F_2 \,.
		\end{eqnarray}
		Indeed, we could have chosen an even character for the superfield by choosing a different expansion for the fields such as 
		\begin{align}\label{AlternativeExpAntiChr}
			Z &= \bar Z_1 + \l^2 \bar Z_2\,, & P_R \chi & = \l^{-1} P_R \chi_1 + \l P_R \chi_2\,, & \bar F & =  \l^{-2} \bar F_1 + F_2 \,,
		\end{align}
		which also preserves the structure of the lowest-order superfield. Nevertheless, the relative expansion character remains unchanged, thus, the transformation rules for the expanded anti-chiral field is independent of the choice such parametrization. We can now repeat the same procedure that we described for the chiral multiplet. The explicit calculation precisely give rise to the transformation rules that we found by complex conjugation \eqref{ExpandedAntiChiral}.
		
		As we expanded the off-shell transformation rules, it is also possible the off-shell chiral multiplet action. The free kinetic Lagrangian for the chiral multiplet is given by
		\begin{eqnarray}
			\mathcal{L} = -  \partial_A Z \partial^A \bar{Z} - \bar\chi P_R \Gamma^A \partial_A P_L \chi + F \bar F \,.
		\end{eqnarray}
		Once again, we can first switch to curved indices by using the vielbein $E_\m{}^A$ and expand the fields with respect to \eqref{ExpansionChiral} and \eqref{AlternativeExpAntiChr}. At the leading order, we have the chiral multiplet action
		\begin{eqnarray}
			\mathcal{L}_{(-2)} &=&  -  \partial_A Z_1 \partial^A \bar{Z}_1 - \bar\chi_1 \Gamma^A \partial_A P_L \chi_1 + F_1 \bar F_1 \,.
		\end{eqnarray}
		At next to leading order, we obtain the action principle for the chiral multiplet of $\cN=1$ coadjoint Poincar\'e algebra
		\begin{eqnarray}
			\cL_{(0)}  &=& -  \partial_A Z_1 \partial^A \bar{Z}_2 -   \partial_A Z_2 \partial^A \bar{Z}_1 - \bar\chi_1 \Gamma^A \partial_A P_L \chi_2 - \bar\chi_2 \Gamma^A \partial_A P_L \chi_1 + F_1 \bar F_2 + F_2 \bar F_1   \,.
		\end{eqnarray}
		Note that the field equations for $Z_2, P_L \chi_2, F_2$ from the variation of $\cL_2$ are consistent with the field equations for $Z_1, P_L \chi_1, F_1$ from the variation of $\cL_1$. This is also a typical property of expanded fields that the higher-order fields arise as Lagrange multiplier to lower-order field equations \cite{Hansen:2020pqs}.


		\section{Three-Dimensional ${g} = (1,1)$ $\cN=2$ Superalgebra}  \label{Section4}
		\paragraph{}
		As a second example, we discuss a non-relativistic algebra presented in \cite{Hansen:2019pkl} in three-dimensions, its $\cN=2$ supersymmetric extension, and its scalar multiplet. As mentioned, certain class of non-relativistic algebras can be obtained by a Lie algebra expansion of a spacetime decomposed Poincar\'e algebra \cite{Bergshoeff:2019ctr}. The bosonic part of the algebra of \cite{Hansen:2019pkl} in $D$-dimensions is one of these algebras that is related to the spacetime decomposition of the three-dimensional Poincar\'e algebra. In this section, we shall focus on $D=3$ and first extend the construction of \cite{Bergshoeff:2019ctr} to include $\cN=2$ supersymmetry. Once the expansion is obtained, we will then follow the prescription for the expansion of the matter multiplets and obtain the $g=(1,1)$ scalar multiplet. 
		
		\subsection{The Expansion of the Spacetime Decomposed Poincar\'e Superalgebra}
		\paragraph{}
		The three-dimensional $\cN=2$ Poincar\'e superalgebra is given by the following non-vanishing (anti-)commutators
		\begin{align}
			[\widehat J_A, \widehat J_B] & = \epsilon_{ABC} \widehat J^C \,, & [\widehat J_A, \widehat P_B] & = \epsilon_{ABC} \widehat P^C \,, & [\widehat J_A, \widehat Q^{1,2}] & =- \frac12 \gamma_A \widehat Q^{1,2} \,, \nonumber\\ 
			\{\widehat Q_\alpha^1 ,\widehat Q_\beta^1 \} & = 2 (\gamma_A C^{-1})_{\alpha \beta} \widehat P^A \,, & \{\widehat Q_\alpha^2 ,\widehat Q_\beta^2 \} & = 2 (\gamma_A C^{-1})_{\alpha \beta} \widehat P^A \,.
		\end{align}
		Here $\widehat P_A$ is the generator of translations, $\widehat J_A$ is for the dual Lorentz transformations and $\widehat Q^{1,2}_\alpha$ are two-component Majorana spinors and they represent the generators of supersymmetry transformations. We may decompose the three-dimensional index $A$ as $A = (0,a)$ and define
		\begin{eqnarray}
			\widehat J_A = (\widehat J, \widehat G_a)\,, \qquad \widehat P_A = (\widehat H, \widehat P_a)\,.
		\end{eqnarray}
		Furthermore, if we introduce \cite{deAzcarraga:2019mdn}
		\begin{eqnarray}
			\widehat Q^\pm = \frac12 \left(Q^1 \pm \g_0 Q^2 \right) \,.
		\end{eqnarray}
		The decomposed algebra is given by \cite{deAzcarraga:2019mdn}
		\begin{align}\label{DecomPoincare}
			[\widehat G_a, \widehat G_b] & = \epsilon_{ab} \widehat J \,, & [\widehat J, \widehat G_a] & = - \epsilon_{ab} \widehat G^b \,,& [\widehat J, \widehat P_a] & = - \epsilon_{ab} \widehat P^b \,, \nonumber\\
			[\widehat G_a, \widehat P_b] & = \epsilon_{ab} \widehat H \,, &  [\widehat G_a, \widehat H] & = - \epsilon_{ab} \widehat P^b \,, & [\widehat J, \widehat Q^\pm] & = - \frac12 \gamma_0 \widehat Q^\pm \,,\nonumber\\
			[\widehat G_a, \widehat Q^\pm] & = - \frac12 \gamma_a \widehat Q^\mp  \,, & \{ \widehat Q^+_\alpha , \widehat Q_\beta^+ \} & = (\gamma_0 C^{-1}) \widehat H \,, & \{ \widehat Q^+_\alpha , \widehat Q_\beta^- \} & = - (\gamma^a C^{-1}) \widehat P_a \,,\nonumber\\
			\{ \widehat Q^-_\alpha , \widehat Q_\beta^- \} & = (\gamma_0 C^{-1}) \widehat H \,,
		\end{align}
		This $[B,F]$ and $\{F,F\}$ commutators of this algebra is of the desired form for the Lie algebra expansion: the susy generators give rise to both spatial translations $\widehat P_a$ and the temporal translations $\widehat H$ \cite{Gomis:2004pw}. Hence the anti-commutator of two supersymmetry generators in the expanded algebra will lead to diffeomorphisms. We, next, split the algebra into even and odd generators such in accordance with multiplication table \eqref{EvenOdd}
		\begin{eqnarray}\label{3dClass}
			V_0 = \{\widehat J, \widehat H, \widehat Q_+\}\,, \qquad V_1 = \{\widehat P_a, \widehat G_a, \widehat Q_-\} \,.
		\end{eqnarray}
		Note that we may also choose all generators to be an element of $V_0$. Although perfectly valid, this choice does not lead to the desired algebra, thus we will not discuss that possibility here. With this characterization of the algebra, the lowest order expansion, $g=(0,0)$ gives rise to $\cN=2$ Galilei superalgebra with no central extension \cite{Andringa:2013mma}
		\begin{align}\label{GalileiSuperalgebra}
			\left[H, G_a\right] &= -\epsilon_{ab} P^b \,,  & \left[J, G_a\right] &= -\epsilon_{ab} G^b \,, & \left[J, P_a\right] &= -\epsilon_{ab} P^b \,, \nonumber \\     
			[J, Q^\pm] &= -\frac12 \gamma_0 Q^\pm \,, & [G_a, Q^+] &= -\frac12 \gamma_a Q^- \,, & 	 \{Q^+_{\alpha},Q^+_{\beta}\} &= (\gamma_0 C^{-1})_{\alpha\beta} H \,,  \nonumber\\
			\{Q^+_{\alpha},Q^-_{\beta}\} &= -(\gamma^a C^{-1})_{\alpha\beta} P_a \,.
		\end{align}
		At the next order, $g=(1,0)$, we obtain the extended Bargmann superalgebra \cite{Bergshoeff:2016lwr,deAzcarraga:2019mdn}
		\begin{align}\label{ExtendedBargmann}
			\left[H, G_a\right] &= -\epsilon_{ab} P^b \,,  & \left[J, G_a\right] &= -\epsilon_{ab} G^b \,, & \left[J, P_a\right] &= -\epsilon_{ab} P^b \,, \nonumber \\     
			\left[G_a, G_b\right] &= \epsilon_{ab} S \,,  & \left[G_a, P_b\right] &= \epsilon_{ab} M \,,& [J, Q^\pm] &= -\frac12 \gamma_0 Q^\pm \,, \nonumber\\
			[J, F^+] & = -\frac12 \gamma_0 F^+ \,, & [G_a, Q^+] &= -\frac12 \gamma_a Q^- \,, & [G_a, Q^-] &= -\frac12 \gamma_a F^+ \,,  \nonumber\\
			[S, Q^+] &= - \frac12 \gamma_0 F^+ \,, & \{Q^+_{\alpha},Q^+_{\beta}\} &= (\gamma_0 C^{-1})_{\alpha\beta} H\,, &\{Q^+_{\alpha},Q^-_{\beta}\} &= -(\gamma^a C^{-1})_{\alpha\beta} P_a \,,\nonumber\\
			\{Q^-_{\alpha},Q^-_{\beta}\} &= (\gamma_0 C^{-1})_{\alpha\beta} M \,, & \{Q^+_{\alpha}, F^+_\beta\} &= (\gamma_0 C^{-1})_{\alpha\beta} M \,.
		\end{align}
		Here, we set 
		\begin{align}
			\widehat P_a^{(1)} & = P_a \,, & \widehat G_a^{(1)} & = G_a \,, & \widehat H^{(0)} & = H \,, & \widehat H^{(2)}  & = M  \,, & \widehat J^{(0)} & = J \,,\nonumber\\
			\widehat J^{(2)}  & = S \,, & \widehat Q^{+(0)} &= Q^+ \,, & \widehat Q^{+(2)} &= F^+\,, & \widehat Q^{-(1)} &= Q^- \,.
		\end{align}
		As we will discuss shortly, this algebra is an example of the consistent truncation with $N_0 = N_1 + 1$. Hence the methodology that we present here does not give rise to supermultiplets of the extended Bargmann superalgebra. 
		
		At order $g=(1,1)$, we obtain the $\cN = 2$ supersymmetric extension of the algebra of \cite{Hansen:2019pkl}, which picks up the following commutation relations in addition to that of \eqref{ExtendedBargmann}
		\begin{align}\label{Newtonian}
			[J, T_a] & = - \e_{ab} T^b \,, & [S, P_a] &= - \e_{ab} T^b \,, & [J, B_a] & = - \e_{ab} B^b \,, \nonumber\\
			[M, G_a] & = - \e_{ab} T^b \,, & [H, B_a] & = - \e_{ab} T^b \,, & [S, G_a] &= - \e_{ab} B^b \,,\nonumber\\
			[J, F^-] & = -\frac12 \gamma_0 F^-  \,, & [G_a, F^+] &= -\frac12 \gamma_a F^- \,, & [B_a, Q^+] &   -\frac12 \gamma_a F^- \,,\nonumber\\
			[S, Q^-] &= - \frac12 \gamma_0 F^- \,, & 	\{Q^+_{\alpha},F^-_{\beta}\} &= -(\gamma^a C^{-1})_{\alpha\beta} T_a \,, &   \{F^+_{\alpha},Q^-_{\beta}\} &= -(\gamma^a C^{-1})_{\alpha\beta} T_a \,, 
		\end{align}
		where we set
		\begin{align}
			\widehat P_a^{(3)} & = T_a \,, & \widehat G_a^{(3)} & = B_a \,,  & \widehat Q^{-(3)} &= F^- \,.
		\end{align}
		With this algebra in hand, we can proceed to the expansion of three-dimensional $\cN=2$ scalar multiplet.
		\subsection{The $g=(1,1)$ Scalar Multiplet}
		\paragraph{}
		The scalar multiplet of three-dimensional $\cN=2$ Poincar\'e superalgebra consists of two real dynamical scalar fields $\varphi_{1,2}$, two Majorana spinors $\chi_{1,2}$ and two real auxiliary scalar fields $F_{1,2}$. The transformation rules for these fields are given by \cite{Bergshoeff:2015ija} 
		\begin{eqnarray}
			\delta \varphi_1 &=& \x^A \partial_A \varphi_1 + \bar\epsilon_1 \chi_1 + \bar\epsilon_2 \chi_2 \,,\nonumber\\
			\delta \varphi_2 &=&  \x^A \partial_A \varphi_1  + \bar\epsilon_1 \chi_2 - \bar\epsilon_2 \chi_1 \,,\nonumber\\
			\delta \chi_1 &=&  \x^A \partial_A \chi_1 + \frac14 \L^{AB} \g_{AB} \chi_1 + \frac14 \gamma^A \partial_A \varphi_1 \epsilon_1 -  \frac14 \gamma^A \partial_A \varphi_2 \epsilon_2 - \frac14 F_1 \epsilon_1 - \frac14 F_2 \epsilon_2  \,,\nn\\
			\delta \chi_2 &=&  \x^A \partial_A \chi_2 + \frac14 \L^{AB} \g_{AB} \chi_2 + \frac14 \gamma^A \partial_A \varphi_1 \epsilon_2 +  \frac14 \gamma^A \partial_A \varphi_2 \epsilon_1 - \frac14 F_2 \epsilon_1 + \frac14 F_2 \epsilon_1  \,,\nn\\
			\delta F_1 &=& \x^A \partial_A F_1 - \bar\epsilon_1 \gamma^A \partial_A \chi_1 + \bar\epsilon_2 \gamma^A \partial_A \chi_2 \,,\nn\\
			\delta F_2 &=& \x^A \partial_A F_2 - \bar\epsilon_1 \gamma^A \partial_A \chi_2 - \bar\epsilon_2 \gamma^A \partial_A \chi_1  \,,
		\end{eqnarray}
		Here, $\e_{1}$ and $\e_{2}$ are the parameters of the supersymmetry generators, $\widehat Q_{1}$ and $\widehat Q_2$, respectively. In order to be able to expand this multiplet in accordance with the spacetime decomposed algebra \eqref{3dClass}, we first split the parameters as
		\begin{align}
			\x^A & = \{\x^0 \equiv \x\,, \x^a\} & \L^{AB} = \{\L^{0a} \equiv \l_G^a \,, \L^{ab} = \e^{ab} \l_J \}
		\end{align}
		and redefine the parameters $\e_{1,2}$ and the spinors $\chi_{1,2}$ as
		\begin{eqnarray}\label{3dSpinor}
			\epsilon_{\pm} = \frac12 (\e_1 \pm \g_0 \e_2) \,, \qquad \text{and} \qquad \chi_\pm = \frac12 (\chi_1 \pm \g_0 \chi_2) \,.
		\end{eqnarray}
		which gives rise to the following definitions for the transformation rules
		\begin{eqnarray}\label{PreExpand}
			\delta \varphi_1 &=& 2 \bar \epsilon_+ \chi_+ + 2 \bar\epsilon_- \chi_- + \xi^a \partial_a \varphi_1+ \xi  \partial_t \varphi_1  \,,\nonumber\\
			\delta \varphi_2 &=& 2 \bar\epsilon_+ \gamma^0 \chi_+ - 2 \bar\epsilon_- \gamma^0 \chi_- + \xi^a \partial_a \varphi_2+ \xi  \partial_t \varphi_2 \,,\nonumber\\
			\delta \chi_+ &=& \frac14 \gamma^0 \partial_t \varphi_1 \e_+ + \frac14 \partial_t \vf_2 \e_+ + \frac14 \g^a \partial_a \vf_1 \e_- - \frac14 \g^{0a} \partial_a \vf_2 \e_- - \frac14 F_1 \e_- - \frac14 F_2 \g_0 \e_-  \nonumber\\
			&&  + \xi^a \partial_a \chi_+  + \xi  \partial_t \chi_+   -\frac12 \lambda_J \g_0 \chi_+ -\frac12 \lambda_G^a \g_a \chi_- \,,\nn\\
			\delta \chi_- &=& \frac14 \gamma^0 \partial_t \vf_1 \e_- + \frac14 \g^a \partial_a \vf_1 \e_+ - \frac14 \partial_t \vf_2 \e_- + \frac14 \g^{0a} \partial_a \vf_2 \e_+ - \frac14 F_1 \e_+ + \frac14 F_2 \g_0 \e_+ \nonumber\\
			&&  + \xi^a \partial_a \chi_- + \xi  \partial_t \chi_-    -\frac12 \lambda_J \g_0 \chi_- -\frac12 \lambda_G^a \g_a \chi_+ \,,\nn\\
			\delta F_1 &=& -2 \bar\e_+ \g^0 \partial_t \chi_- - 2 \bar\e_- \g^0 \partial_t \chi_+ - 2 \bar\e_+ \g^a \partial_a \chi_+ - 2 \bar\e_- \g^a \partial_a \chi_-  + \xi^a \partial_a F_1 + \xi  \partial_t F_1 \,,\nn\\
			\delta F_2 &=& - 2 \bar\e_+ \partial_t \chi_- + 2 \bar\e_- \partial_t \chi_+ - 2 \bar\e_+ \g^{a0} \partial_a \chi_+ + 2 \bar\e_- \g^{a0} \partial_a \chi_-   + \xi^a \partial_a F_2 + \xi  \partial_t F_2 \,,
		\end{eqnarray}
		where $\partial_t$ represents the time derivative $\partial_t  \equiv \partial_0$. This multiplet can now be expanded in by using the commutator structure \eqref{DD3} as well as the leading order behavior of the spatial dreibein $\e_\m{}^a$ and the temporal dreibein $\tau_\m$, i.e.
		\begin{eqnarray}
			e_\m{}^a \quad \to \quad \l e_\m{}^a \,, \qquad \qquad \tau_\m \quad \to \quad \tau_\m \,.
		\end{eqnarray}
		Note that both of the dreibein are necessary since a curved index $\mu$ is expanded as 
		\begin{eqnarray}
			A_\m = e_\m{}^a A_a + \tau_\m A_0 \,, \qquad \Rightarrow \qquad A_a = e^\m{}_a A_\m \,, \quad \text{and} \quad A_0 = - \tau^\m A_\m \,,
		\end{eqnarray}
		where we defined the inverse spatial ($e^\m{}_a$) and temporal ($\t^\m$) dreibein in accordance with the relations
		\begin{eqnarray}
			e_\m{}^a e^{\m}{}_b = \d^a{}_b \,, \qquad \text{and} \qquad \t^\m \t_\m =  - 1\,.
		\end{eqnarray}
		Let us first expand this algebra to $g=(1,0)$ order to see why such an expansion fails. Unlike the coadjoint expansion, the lowest order algebra is different than the algebra that we expand due to the $\{Q^-, Q^-\}$ anti-commutator. However, the rest of the algebra stays intact. Thus, ignoring the $Q^-$ transformations, we may still find out the expansion character of the fields from the transformation rules \eqref{PreExpand}. As only the relatives character matters, we can give the dynamical scalars an even expansion character, say they are expanded as $\vf_1 = \phi_1 + \l^2 \phi_2 \,,$ and $\vf_2  = \phi_3 + \l^2 \phi_4 $. The $Q^+$ transformation of $\vf_1$ or $\vf_2$ fixes the character of $\chi_+$ to be the same as the scalar fields. On the other hand, the $Q^+$ transformation of $\chi_-, F_1$ and $F_2$ indicate that they must have an odd character and their expansion should start with $\lambda^{-1}$-order
		\begin{align}
			\vf_1 & = \phi_1 + \l^2 \phi_2 \,,& \vf_2 & = \phi_3 + \l^2 \phi_4 \,, & \chi_+ & = \psi_1 + \l^2 \psi_2 \,,\nonumber\\
			\chi_- & = \frac1\l  \psi_3 \,, &  F_1 & = \frac1\l  S_1 \,,& F_2 & = \frac1\l  S_3  \,. 
		\end{align}
		As the expansion is of order $g=(1,0)$, we expand the even fields for two orders but the odd ones for one order. In this case, it is sufficient to consider the transformation rules of $\f_2, \p_1$ and $\p_2$ to show that the $g=(1,0)$ expansion is not closed on the extended Bargmann superalgebra \eqref{ExtendedBargmann} 
		\begin{eqnarray}
			\d \phi_2 &=&  2 \bar\e_+ \psi_2+2 \bar\eta_+ \psi_1 + \xi^a \partial_a \phi_2+ \xi  \partial_t \phi_2 + \L_M \partial_t \f_1   \,,\nonumber\\
			\delta \p_1 &=&  \frac14  \g^0\e_+ \partial_t \f_1 +\frac14  \g_a \e_- \partial_a \f_1  -\frac14  \g^{0a}\e_- \partial_a \f_3 + \frac14 S_3 \g_0 \e_- - \frac14 S_1  \e_- + \frac14 \partial_t \f_3 \e_+  \nonumber\\ 
			&& - \frac12 \l_G^a \g_a \p_3- \frac12 \l  \g_0 \p_1 + \xi^a \partial_a \p_1 + \xi  \partial_t \p_1 \,,\nonumber\\
			\delta \p_2 &=& \frac14 \g^0\e_+ \partial_t \f_2+\frac14 \g^0\eta_+ \partial_t \f_1+\frac14 \g^a \e_- \partial_a \f_2    -\frac14 \g^{0a}\e_- \partial_a \f_4- \frac14 S_2\e_-  +\frac14 \partial_t\f_4 \e_+ \nonumber\\
			&& +\frac14 \partial_t\f_3 \eta_+ 	-\frac12 \l_J \g_0 \p_2	-\frac12 \l_S \g_0 \p_1 + \xi^a \partial_a \p_2+ \xi  \partial_t \p_2+\L_M \partial_t \p_1 	\,.
		\end{eqnarray}
		Here, $\x^a$ are the parameter for spatial translations, $\x$ for temporal translations and $\L_X$ for $X$-generators with $X = \{J, S, M, G_a\}$. Furthermore, $\e_\pm$ correspond to the parameters for $Q^\pm$ and $\eta_+$ is for $F^+$. It is evident with these transformation rules that the commutator $\{Q^-, Q^-\} \sim M$ does not close on the field $\phi_2$ since it is inert under $Q^-$ transformation, yet transforms non-trivially under $M$. Nevertheless, since $M$ is central in the extended Bargmann superalgebra \eqref{ExtendedBargmann}, it is tempting to ignore the $M$ transformations of the fields. In that case, however, the commutator $\{Q^+, F^+\} \sim M$ fails to close off-shell on $\f_2$. The root cause of this problem goes back to the expansion of the Jacobi identites \eqref{ExpandedJacobi}, in particular, to the expansion of 
		\begin{eqnarray}
			[\e_1^\a Y_\a, [\e_2^\b Y_\b, \vf_1]] - 	[\e_2^\b Y_\b, [\e_1^\a Y_\a, \vf_1]] & = [[\e_1^\a Y_\a, \e_2^\b Y_\b], \vf_1] \,.
		\end{eqnarray}
		Because of the fact that the structure constants do not get modified in the Lie algebra expansion, the expansion of the Jacobi identities \eqref{ExpandedJacobi} must either be preserved or vanish identically, that is, if one of the non-zero commutators become zero in the expansion, all other commutators must follow. In the case of the scalar multiplet, highest order even fields such as $\f_2, \f_4$ and $\p_2$, have cannot non-trivial $Q^-$ transformation because there the odd fields are not expanded to include $\mathcal{O}(\l)$ terms. Thus  $\{Q^-, Q^-\} \sim M$ is failed on these fields. This, as a matter of fact, is a typical property of $g=(N_1+1, N_1)$ type of expansion. The generator of time translations $H$ only appears in the following commutation relations in spacetime decomposed Poincar\'e algebra \eqref{DecomPoincare} has the structure
		\begin{align}
			[\widehat G_a, \widehat H] & = - \epsilon_{ab} \widehat P^b\,,  & \{ \widehat Q^+_\alpha , \widehat Q_\beta^+ \} & = (\gamma_0 C^{-1}) \widehat H \,, & \{ \widehat Q^-_\alpha , \widehat Q_\beta^- \} & = (\gamma_0 C^{-1}) \widehat H \,,
		\end{align}
		The first commutator implies that as long as we consider $g=(N_1+1,N_1)$-type of expansion, the highest order generator in the expansion of $\widehat H$ is always central in the bosonic part of the algebra. On the other hand, this highest order generator also shows up on the right-hand side in the expansion of the second and the third anti-commutator. In particular, the anti-commutator of the highest and the lowest generators in the expansion of $Q^-$ gives rise to the highest order generator in the expansion of $\widehat H$. On the other hand, because of the $\l^{-1}$ relative expansion character of the odd-fields, which we found that applies to the vast majority of matter multiplets including scalar, chiral, vector, real, and linear multiplets, the highest order even fields are always inert under the hugest order generators in the expansion of $Q^-$ and yet they transform non-trivially under the highest order generator in the expansion of $\widehat H$. Consequently, $g=(N_1+1,N_1)$-type of expansion always fail on the highest even fields.
		
		Next, we proceed to the scalar multiplet of $g=(1,1)$ superalgebra. In this case, the expansion of the fields are realized as
		\begin{align}
			\vf_1 & = \phi_1 + \l^2 \phi_2 \,,& \vf_2 & = \phi_3 + \l^2 \phi_4 \,, & \chi_+ & = \psi_1 + \l^2 \psi_2 \,,\nonumber\\
			\chi_- & = \frac1\l  \psi_3  + \l \p_4 \,, &  F_1 & = \frac1\l  S_1 + \l S_2 \,,& F_2 & = \frac1\l  S_3 + \l S_4 \,. 
		\end{align}
		The odd fields now pick up the necessary $\mathcal{O}(\l)$ terms and the transformation rules are given by
		\begin{eqnarray}
			\d \phi_1 &=& 2 \bar\e_- \psi_3 +2 \bar\e_+ \psi_1 + \xi^a \partial_a \phi_1+ \xi  \partial_t \phi_1 \,,\nonumber \\
			\d \phi_2 &=& 2 \bar\eta_- \psi_3 + 2 \bar\e_- \psi_4 + 2 \bar\e_+ \psi_2+2 \bar\eta_+ \psi_1 + \xi^a \partial_a \phi_2+ \xi  \partial_t \phi_2  + \l^a_T \partial_a \f_1 + \L_M \partial_t \f_1  \,,\nonumber\\
			\d \phi_3 &=& -2 \bar\e_- \g^0\psi_3 +2 \bar\e_+\g^0 \psi_1 + \xi^a \partial_a \phi_3+ \xi  \partial_t \phi_3 \,,\nonumber \\
			\d \phi_4 &=& -2 \bar\eta_-\g^0 \psi_3 - 2 \bar\e_-\g^0 \psi_4 + 2 \bar\e_+\g^0 \psi_2+2 \bar\eta_+\g^0 \psi_1 + \xi^a \partial_a \phi_4+ \xi  \partial_t \phi_4 \nn\\
			&& + \l^a_T \partial_a \f_3 + \L_M \partial_t \f_3  \,,\nonumber\\
			\delta \p_1 &=&  \frac14  \g^0\e_+ \partial_t \f_1 +\frac14  \g^a \e_- \partial_a \f_1  -\frac14  \g^{0a}\e_- \partial_a \f_3 + \frac14 S_3 \g_0 \e_- - \frac14 S_1  \e_- + \frac14 \partial_t \f_3 \e_+  \nonumber\\ 
			&& - \frac12 \l_G^a \g_a \p_3- \frac12 \l_J  \g_0 \p_1 + \xi^a \partial_a \p_1 + \xi  \partial_t \p_1 \,,\nonumber\\
			\delta \p_2 &=& \frac14 S_4 \g_0 \e_- +\frac14 S_3 \g_0 \eta_- +\frac14 \g^0\e_+ \partial_t \f_2+\frac14 \g^0\eta_+ \partial_t \f_1+\frac14 \g^a\e_- \partial_a \f_2 +\frac14 \g^a\eta_- \partial_a \f_1  \nonumber\\
			&& -\frac14 \g^{0a}\e_- \partial_a \f_4-\frac14 \g^{0a}\eta_- \partial_a \f_3- \frac14 S_2\e_- - \frac14 S_1\eta_-+\frac14 \partial_t\f_4 \e_+ +\frac14 \partial_t\f_3 \eta_+	 	\nn \\	
			&& 	-\frac12 \l_J \g_0 \p_2	-\frac12 \l_S \g_0 \p_1 -\frac12 \l_G^a \g_a \p_4-\frac12 \l_B^a \g_a \p_3 + \xi^a \partial_a \p_2+ \xi  \partial_t \p_2\nonumber\\
			&& +\l_T^a \partial_a \p_1 +\L_M \partial_t \p_1  \,,	\nonumber	\\
			\delta \psi_3 &=&  \frac14 S_3 \g_0\e_+ +\frac14 \g^a\e_+ \partial_a \f_1 +\frac14 \g^{0a}\e_+ \partial_a \f_3-\frac14 S_1 \e_+ + \xi^a \partial_a \p_3+ \xi  \partial_t \p_3 -\frac12 \l_J \g_0 \p_3 \,,\nonumber\\
			\d \p_4 &=& \frac14 S_4 \g_0 \e_+ +\frac14 S_3 \g_0 \eta_+ +\frac14 \g^0\e_- \partial_t \f_1+\frac14 \g^a\e_+ \partial_a \f_2+\frac14 \g^a\eta_+ \partial_a \f_1 +\frac14 \g^{0a}\e_+ \partial_a \f_4  \nonumber\\
			&& +\frac14 \g^{0a}\eta_+ \partial_a \f_3-\frac14  \partial_t \f_3 \e_- - \frac14 S_2\e_+ - \frac14 S_1\eta_+ -\frac12 \l_J \g_0 \p_4	\, \nonumber\\
			&& 	-\frac12 \l_S \g_0 \p_3 -\frac12 \l_g^a \g_a \p_1+\xi^a \partial_a \p_4+ \xi  \partial_t \p_4+\l_T^a \partial_a \p_3 +\L_M \partial_t \p_3  \,,	\nonumber\\
			\d S_1 &=& -2 \bar\e_- \g^a \partial_a \p_3 - 2 \bar\e_+ \g^0 \partial_t \p_3 - 2 \bar\e_+ \g^a \partial_a \p_1+ \xi^a \partial_a S_1+ \xi  \partial_t S_1 \,,\nonumber\\
			\d S_2 &=& -2 \bar\e_- \g^0 \partial_t \p_1 - 2 \bar\e_- \g^a \partial_a \p_4 - 2 \bar\eta_- \g^a \partial_a \p_3-2 \bar\e_+ \g^0 \partial_t \p_4 - 2 \bar\e_+ \g^a \partial_a \p_2  \nonumber\\
			&& - 2 \bar\eta_+ \g^0 \partial_t \p_3  - 2 \bar\eta_+ \g^a \partial_a \p_1+ \xi^a \partial_a S_2+ \xi  \partial_t S_2+\l_T^a \partial_a S_1+\L_M \partial_t S_1 \,,\nonumber\\
			\d S_3 &=& -2 \bar\e_+  \partial_t \p_3 - 2 \bar\e_- \g^{0a} \partial_a \p_3 + 2 \bar\e_+ \g^{0a} \partial_a \p_1+ \xi^a \partial_a S_3+ \xi  \partial_t S_3 \,,\nonumber\\
			\d S_4 &=& 2 \bar\e_- \partial_t \p_1 - 2 \bar\e_+  \partial_t \p_4 - 2 \bar\eta_+  \partial_t \p_3-2 \bar\e_- \g^{0a} \partial_a \p_4 - 2 \bar\eta_- \g^{0a} \partial_a \p_3  \nonumber\\
			&& + 2 \bar\e_+ \g^{0a} \partial_a \p_2  + 2 \bar\eta_+ \g^{0a} \partial_a \p_1+ \xi^a \partial_a S_4+ \xi  \partial_t S_4+\l_T^a \partial_a S_3+\L_M \partial_t S_3 \,.
		\end{eqnarray}
		Here, the new parameters $\L_T^a, \L_B^a, \eta_-$ are the parameters of $T^a, B^a$ and $F^-$ transformations, respectively. We  explicitly checked by using \textit{GammaMaP} \cite{Kuusela:2019iok} that this multiplet is indeed closed off-shell given that the third parameters of the transformation rules are given by
		\begin{align}
			\{Q^+_{\alpha},Q^+_{\beta}\} &= (\gamma_0 C^{-1})_{\alpha\beta} H \,,  & \Rightarrow & & \xi_3&=\bar\epsilon_{2+}\gamma_0C^{-1}\epsilon_{1+}  \,,\nn \\
			\{Q^+_{\alpha},Q^-_{\beta}\} &= -(\gamma^a C^{-1})_{\alpha\beta} P_a \,,\quad & \Rightarrow & & \xi^a_3 &=- \bar\epsilon_{+}\gamma^aC^{-1}\epsilon_{-} \,, \nn \\
			\{Q^-_{\alpha},Q^-_{\beta}\} &= (\gamma_0 C^{-1})_{\alpha\beta} M \,, & \Rightarrow & & (\Lambda_M)_3 &=\bar\epsilon_{1-}\gamma_0C^{-1}\epsilon_{2-} \,, \nn \\ 
			\{Q^+_{\alpha}, F^+_\beta\} &= (\gamma_0 C^{-1})_{\alpha\beta} M  \,,& \Rightarrow & & (\Lambda_M)_3 &=\bar\epsilon_{+}\gamma_0C^{-1}\eta_{+} \,, \nn \\ 
			\{Q^+_{\alpha},F^-_{\beta}\} &= -(\gamma^a C^{-1})_{\alpha\beta} T_a \,,  & \Rightarrow & & (\lambda_T)_3^a &=-\bar\epsilon_{+}\gamma^aC^{-1}\eta_{-} \,, \nn \\  
			\{F^+_{\alpha},Q^-_{\beta}\} &= -(\gamma^a C^{-1})_{\alpha\beta} T_a \,, & \Rightarrow & & (\lambda_T)_3^a &=-\bar\eta_{+}\gamma^aC^{-1}\epsilon_{-}\,.
		\end{align}
		Finally, as the action principle for the $\cN=2$ scalar multiplet is given by \cite{Alkac:2014hwa} 
		\begin{eqnarray}
			\cL = - \partial_A \vf_1 \partial^A \vf_1 - \partial_A \vf_2 \partial^A \vf_2- 4 \bar\chi_1 \gamma^A \partial_A \chi_1 -  4 \bar\chi_2 \gamma^A \partial_A \chi_2 + F_1^2 + F_2^2 \,,
		\end{eqnarray}
		we can implement the spacetime decomposition in accordance with \eqref{3dSpinor} which gives rise to the action principle that we can use to expand to find order $g=(N_0,N_0)$ extended algebras
		\begin{eqnarray}
			\cL &=& \partial_t \vf_1 \partial_t \vf_1 - \partial^a \vf_1 \partial_a \vf_1  +  \partial_t \vf_2 \partial_t \vf_2 - \partial^a \vf_2 \partial_a \vf_2  - 8 \bar\chi_+ \gamma^0 \partial_t \chi_+ \nonumber\\
			&& - 8 \bar\chi_- \gamma^0 \partial_t \chi_-  - 8 \bar\chi_+ \g^a \partial_a \chi_- - 8 \bar\chi_- \g^a \partial_a \chi_+ + F_1^2 + F_2^2 \,.
		\end{eqnarray}
		At order $g=(0,0)$, we have the same action as one would get from contraction. As we have real scalar fields, it is necessary to keep in mind that after the contraction, the dynamical kinetic scalar has no longer a time derivative in the field equations due to the appearence of the factor $1/c$, i.e. Klein-Gordon equation for a real scalar $\f$ field has the following non-relativistic limit \cite{Bergshoeff:2017vjg}
		\begin{eqnarray}
			\frac{1}{c^2} \frac{\partial}{\partial t} \phi - \nabla^2 \phi + \frac{m^2 c^2}{\hbar^2} \phi = 0 \,, \qquad \xRightarrow{c \to \infty}  \qquad - \nabla^2 \phi + \frac{m^2 c^2}{\hbar^2} \phi = 0
		\end{eqnarray}
		where we keep $mc/\hbar$ fixed. Thus, the lowest order action, which is invariant under $\cN=2$ Galilei superalgebra, has no time derivative for the scalar fields, and it is given by $\mathcal{O}(\l^{-2})$ action
		\begin{eqnarray}
			\cL_{(-2)} &=& -  \partial^a \f_1 \partial_a \f_1 -  \partial^a \f_3 \partial_a \f_3 - 8 \bar\p_3 \g^0 \partial_t \p_3 - 8 \bar\p_1 \g^a \partial_a \p_3 - 8 \bar\p_3 \g^a \partial_a \p_1 + S_1^2 + S_3^2 \,. \qquad 
		\end{eqnarray} 
		At order $g=(1,1)$, the invariant action is therefore given by the $\mathcal{O}{(\l^0)}$ term in the expansion 
		\begin{eqnarray}
			\cL_{(0)} &=& \partial_t \f_1 \partial_t \f_1 + \partial_t \f_3 \partial_t \f_3 - \partial^a \f_1 \partial_a \f_2 -  \partial^a \f_3 \partial_a \f_4 - 8 \bar\p_1 \g^0 \partial_t \p_1 - 8 \bar\p_3 \g^0 \partial_t \p_4 \nonumber\\
			&&   - 8 \bar\p_4 \g^0 \partial_t \p_3 -8 \bar\p_1 \g^a \partial_a \p_4 - 8 \bar\p_2 \g^a \partial_a \p_3 -8 \bar\p_4 \g^a \partial_a \p_1  -8 \bar\p_3 \g^a \partial_a \p_2 \nonumber\\
			&& + S_1 S_2 + S_3 S_4 \,. 
		\end{eqnarray}
		As in the case of chiral multiplet for the $\cN=1$ coadjoint Poincar\'e supersymmetry, the higher-order fields such as $\f_2, \f_4, \p_2, \p_4, S_2$ and $S_4$ appear as Lagrange multiplier that imposes the lowest order field equations given by the field equation of the Lagrangian $\cL_{(-2)}$. This point is also relevant to the reducible structure of the multiplets, so let us be more precise. Consider the variation of the action
		\begin{eqnarray}
			\delta S =	\int  \left( \text{Lower-order field equation} \right) \delta(\text{higher-order field}) + \ldots  \,,
		\end{eqnarray}
		where the ellipses represent the variation of lower-order fields. This structure of the variation has two important consequences. First, as the lower-order equations are covariant under lower-order transformations, the higher-order transformation rules in the higher-order fields must have the same differential structure as their ancestors. The only difference is that the operators also have a matrix structure, taking the higher-order fields to lower-order ones. Second, as the lower-order field equations are preserved, one can always consistently truncate lower-order fields in the multiplet, thus the reducible structure. When this is done, the higher-order fields become inert under higher-order transformations and become identical to the lower-order ones.
		
		\section{Discussion}\label{Comments}
		\paragraph{}
		In this paper, we present a methodology to generate the matter multiplets of non-relativistic algebras and coadjoint Poincar\'e algebra starting from the matter multiplets of the Poincar\'e algebra itself, or its spacetime decomposition. The methodology relies on the fact that the rigid limit of the transformation rules of local, off-shell matter multiplet on a flat background can be performed by setting all fields of the off-shell supergravity multiplet to zero, except for the vierbein and temporal vierbein. This fact allows us to develop an auxiliary scheme that we only use these two fields, or their inverse, in the off-shell local description of the multiplets to relate the flat indices to curved indices. Once this is achieved, we were able to expand the commutator of covariant derivatives, $[\cD_\m, \cD_\n] = - \d(R_{\m\n})$, to determine the transformation rules for matter multiplets of larger algebras by using the Lie algebra expansion procedure.
		
		As one of the key points of our procedure is that the truncation must occur at order $g=(N_1,N_1)$, it is not obvious how to use our methodology to find out the matter multiplets of algebras of order $g=(N_1+1,N_1)$, such as the extended Bargmann algebra. One candidate could be that these algebras could appear as the order $g=(N_1,N_1)$ algebras of some other smaller algebra(s). For instance, in Ref.\cite{Penafiel:2019czp}, it was shown that the extended Bargmann algebra is actually of order $g=(1,1)$ when the Nappi-Witten algebra is expanded instead of a spacetime decomposed Poincar\'e superalgebra. One may hope that such smaller algebras can be obtained as a subalgebra of the Poincar\'e superalgebra in one higher dimension. In the case of the bosonic part of the three-dimensional Nappi-Witten algebra, it is indeed the case. However, it is not obvious if the supersymmetric part of the algebra follows. Furthermore, when the light-cone decomposition is performed, one needs to mix the generators to obtain the desired commutation relations for the non-relativistic algebra. This would alter the physical definitions of the generators since they are deformed or completely changed. For instance, in the case of the non-relativistic conformal extension of the extended Bargmann superalgebra, which is a non-trivial extension of the Schr\"odinger supersymmetry \cite{Horvathy:1992pcm,Duval:1993hs,Duval:1995dq}, the small algebra that one expands have the following definition for rotations \cite{Kasikci:2020qsj}
		\begin{eqnarray}
			j =& \frac{1}{6} P_- + \frac{1}{6} K_+ - \frac{2}{3} \widehat J  \,,
		\end{eqnarray}
		where $\mathcal{O}_\pm$ are the light-cone decomposition of operators, $P_A$ is the generators of four-dimensional translations, $K_A$ is for four-dimensional special conformal transformations and $\widehat J_{AB}$ is associated with the four-dimensional Lorentz transformations. These identifications are very different from the spacetime decomposition where the lowest order generators are associated with their standard differential operator definitions. In the case of light-cone decomposition, it is no longer obvious if they make any physical sense, or even the supersymmetry generators mean the usual supersymmetry in the sense that a sequential application of SUSY operators generates diffeomorphisms.
		
		The multiplets that we present here are reducible by the nature of Lie algebra expansion. At this stage, we do not know if there can be irreducible ones. Nevertheless, this seems to be an important next step in understanding the multiplet structure of non-relativistic supersymmetry. Another natural question is that if we can expand local, off-shell matter multiplets and actions as we did here for the global ones. As mentioned, this would require a procedure that can give rise to off-shell non-relativistic supergravity to yield the necessary off-shell background for the matter multiplets. We hope to address these questions in near future.
		
		\section*{Acknowledgment}
		\paragraph{}
		We thank Utku Zorba for various comments and collaboration at the early stages of this work. The work of O.K. is supported by TUBITAK grant 118F091. M.O. is supported in part by TUBITAK grant 118F091, TUBITAK grant 121F064 and Istanbul Technical University Research Fund under grant number TGA-2020-42570. M.O. acknowledges the support by the Distinguished Young Scientist Award BAGEP of the Science Academy. M.O. also acknowledges the support by the Outstanding Young Scientist Award of the Turkish Academy of Sciences (TUBA-GEBIP).

		\providecommand{\href}[2]{#2}\begingroup\raggedright\endgroup

	\end{document}